\documentclass[pre,showpacs,showkeys]{revtex4}

\usepackage{graphicx}
\usepackage{dcolumn}
\usepackage{bm}
\usepackage{subfigure}
\usepackage{amssymb}
\usepackage{graphics}
\usepackage{hyperref}
\usepackage{epsfig}
\usepackage{threeparttable}
\newcommand{\thickhline}{\noalign{\hrule height 0.8pt}}

\begin{document}


\title{Nonmagnetic impurities and roughness effects on the finite temperature magnetic properties of core-shell spherical nanoparticles}
\author{E. Vatansever}\email{erol.vatansever@deu.edu.tr } 
\author{Y. Y\"{u}ksel}
\affiliation{ Department of Physics, Dokuz Eyl\"{u}l University,
Tinaztepe Campus, TR-35160 Izmir, Turkey}
\date{\today}

\begin{abstract}
Being inspired by a recent study [V. Dimitriadis et al. Phys. Rev. B \textbf{92}, 064420 (2015)], 
we study the finite temperature magnetic properties of the spherical nanoparticles with core-shell 
structure including quenched (i) surface and (ii) interface
nonmagnetic impurities (static holes) as well as (iii) roughened interface effects. 
The particle core is composed of ferromagnetic spins, and it is surrounded by a ferromagnetic shell. By means of Monte Carlo simulation
based on an improved Metropolis algorithm, we implement the nanoparticles using classical
Heisenberg Hamiltonians. Particular attention has also been devoted to  elucidate the effects of the particle
size on the thermal and magnetic phase transition  features of these systems. For nanoparticles with
imperfect surface layers, it is found that  bigger particles exhibit lower compensation
point which decreases gradually with increasing amount of vacancies, and vanishes at a
critical value. In view of nanoparticles with diluted 
interface, our Monte Carlo simulation results suggest that there exists a 
region in the disorder spectrum where compensation temperature linearly decreases with 
decreasing dilution parameter. For nanoparticles with roughened  interface, it is observed that  the degree of  
roughness does not play any  significant role on the variation of both the compensation  
point and critical temperature. However,   the low temperature saturation magnetizations of the 
core and shell interface regions sensitively depend on the roughness parameter.
\end{abstract}


\maketitle

\section{Introduction}\label{Intro}
When the size of a magnetic system is reduced to a characteristic length,
the system has a bigger surface to volume ratio giving rise to a great many
outstanding thermal and magnetic properties compared to the conventional
bulk systems \cite{Berkowitz1}. Advanced functional magnetic nanostructures in different geometries,
such as nanowires, nanotubes, nanospheres, nanocubes are center of interest
because of their technological \cite{Hayashi, Guo} and  scientific importance as
well as biomedical  applications \cite{Pankhurst, McNamara, Ivanov}.
From the experimental point of view, many studies have been carried out to discuss
and understand the origin of the fascinating  physical properties observed in
magnetic nanoparticles \cite{Ivanov, Bhatt, Martinez, Estrader, Bhowmik}.   For example, recently the multi-functional
core-shell nanowires have been synthesized by a facile low-cost fabrication
process \cite{Ivanov}. Based on this study, it has been shown that  a multidomain state at
remanence can be  obtained, which is an attractive feature for the
biomedical applications. In another interesting study, the authors show the presence of a robust
antiferromagnetic coupling between core and shell in ferrimagnetic
soft/hard and  hard/soft core-shell nanoparticles based
on Fe-oxides and Mn-oxides \cite{Estrader}. They have also used a  computational
model to support the physical facts observed in the experiment.
Moreover, it is a fact  that core-shell nanoparticle systems exhibit
two important  phenomena, namely  exchange bias  and magnetic proximity effects.
These are completely due to the interface effects of the system. For detailed reviews on the exchange
bias and magnetic proximity phenomena, the readers may follow the
references \cite{Manna, Kiwi, Nogues, Berkowitz2, Iglesias}.

Ferrimagnetic materials have a compensation temperature under certain
conditions. At this special temperature region, the net magnetization of the sample
vanishes below its critical temperature \cite{Neel}. The
phenomenon of ferrimagnetism in bulk material is
associated  with the counteraction of  opposite magnetic moments with unequal magnitudes
located  on different sublattices in the same system. According to the
Refs. \cite{Hansen, Buendia}, interestingly coercive field presents a behavior with a rapid
increment at the compensation point.  Existence of such a point has a technological
importance  \cite{Shieh, Mansuripur}, because at this point only a small
magnetic field is required and  enough to change the sign of the net
magnetization. However,  the origin of the
compensation point found in the nanostructures is quite different
from those observed in the ferrimagnetic bulk materials.   Magnetic nanoparticles
can exhibit a compensation point due to the existence of an
antiferromagnetic interface  coupling at the ferromagnetic core and ferromagnetic shell interface even
if the lattice sites in the  core and shell  parts of the system are occupied by identical
atomic spin moments. Hence, investigation of ferrimagnetism in nanoparticle systems has opened a new and
an intensive field in the research of the critical phenomena
in magnetic nanoparticles. For example, the critical and compensation temperatures
properties of cylindrical nanowire and  nanotube systems have been performed
by means of Effective-Field Theory with single-site correlations \cite{Kaneyoshi1, Kaneyoshi2}.
In these studies, the authors have also focused their attention on the effects of
the surface and its dilution on the magnetic  properties of the considered system, and it is reported that
these systems display a compensation point for appropriate
values of the system parameters. Very recently, thermal and magnetic phase
transition features of a core-shell spherical nanoparticle with binary
alloy shell  have been studied by making use of Monte Carlo simulation based on
single-spin flip Metropolis algorithm \cite{Zaim}. Here, the authors claim
that the system may demonstrate one, two or even three compensation
points depending on the selected Hamiltonian  as well as on the
concentration parameters. In addition to these,
critical behaviors of core-shell nanoparticles with  ferromagnetic materials
but with antiferromagnetic  interface exchange  coupling are studied by means of a self-consistent
local mean-field analysis \cite{Anderson}. It has been found that
compensation temperature  depends on all the  material parameters,
namely the core and shell radius, and the magnetic field.

Although the mechanism and physics underlying of the critical behavior of the magnetic
nanoparticles may be treated and understood with idealized interfaces and
surfaces of the nanoparticle, real magnetic nanoparticles have some small defects.
From this point of view, experimental systems showing exchange bias may contain
statistical  distributions due to the presence of  randomly located defects
in the system \cite{Evans1, Evans2}. Recently, Ho and co-workers have
attempted to address the magnetic properties of a ferromagnetic/antiferromagnetic core-shell
nanospherical particle including the vacancies at the antiferromagnetic interface,
based on Monte-Carlo simulation method \cite{Ho}. It is found that the frustrated spins
at the ferromagnetic interface is another pinning-source generating exchange
bias phenomenon, in addition to the antiferromagnetic shell spins.
Furthermore, the influences of non-magnetic defects on the exchange bias of core-shell nanoparticles
have been  analyzed by benefiting from Monte Carlo simulation, and
it is shown that exchange bias can be tuned by defects in different
positions \cite{Mao}. Apart from these, Evans et al. \cite{Evans1} presented
exchange-bias  calculations for FM core/AFM shell nanoparticles with roughened interfaces.
They showed that the magnitude of exchange bias is strongly correlated with the
degree of roughness. Moreover, in a very recent paper, Dimitriadis et
al. \cite{Dimitriadis} simulated cubic and spherical particles showing exchange bias
phenomenon. According to their results, in terms of exchange bias characters,
the distinction between cubic and spherical particles is lost
for moderate roughness.

Based on the previously published studies, it is possible to mention that
thermal and magnetic properties of the core-shell nanoparticles containing
the surface and interface defects and also roughened interfaces are
complicated and interesting compared to the clean nanoparticle systems.
It is certain that the studies taking these effects into account
play a crucial role in having a
better insight of the physics behind real magnetic nanoparticle systems.
However, much less attention has been given to determine the
influences of the disorder and roughness on the critical behavior of
the core-shell nanoparticles, and  there are still many unresolved issues.
Motivated by these facts, we intend to search  answers for the following
questions:
\begin{itemize}
\item What are the effects of the nonmagnetic impurities  at the surface and
interface of a core-shell type spherical nanoparticle on the
critical and compensation behavior ? Is it possible to control 
the magnetic behavior of the core-shell nanoparticles by means of concentration of the vacancies ?
\item What kind of physical relationships may emerge between the physical properties of the
system and the interface roughness ?
\end{itemize}

The main motivation of the paper is to make an attempt to determine the physical facts underlying these
questions. Furthermore, particular attention has been dedicated to elucidate the effects of the system size
on the critical behavior of the system.  We believe that the
findings obtained in this work would be beneficial for the future theoretical and
experimental research in magnetic nanoparticles including disorder effects.

The outline of the remainder parts of the paper is as follows: In
section \ref{formulation}, we present the model and simulation details. The results
and discussion are given in section \ref{results}, and finally
section \ref{conclude} includes our conclusions.

\section{Model and Simulation Details}\label{formulation}
\begin{figure}
\includegraphics[width=6.0cm]{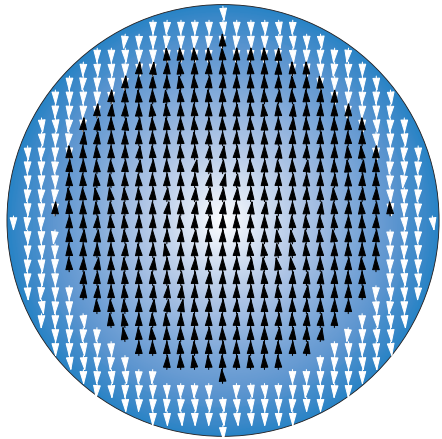}\\
\caption{Schematic representation of an ideally purified spherical nanoparticle composed
of classical Heisenberg spins. FM core of radius $R_{c}$ is coated by an FM shell of
thickness $R_{S}$. Total radius of the particle is denoted by $R=R_{c}+R_{S}$.
At the interface, the outermost core and the innermost shell layers interact
with an AFM exchange coupling.
}\label{fig1}
\end{figure}

We consider a spherical nanoparticle which has been schematically depicted in Fig. \ref{fig1} with a FM core
which is surrounded by a FM shell. At the core-shell interface (which is composed of the outermost core and the
innermost shell layers) we define an AFM exchange coupling. The interface region consists of two successive monolayers.
The total radius of the particle and the thickness of the shell are denoted by $R$ and $R_{S}$, respectively. A classical
Heisenberg spin resides at each lattice site of a simple cubic structure, and the nearest neighbor sites are separated
from each other by unitary lattice spacing. The system can be defined according to the following Hamiltonian
\begin{equation}\label{eq1}
\mathcal{H}=-J_{C}\sum_{<ij>}\textbf{S}_{i}.\textbf{S}_{j}-J_{S}\sum_{<kl>}\textbf{S}_{k}.
\textbf{S}_{l}-J_{IF}\sum_{<jk>}\textbf{S}_{j}.\textbf{S}_{k}
-K_{C}\sum_{i}(S_{i}^{z})^{2}-K_{S}\sum_{k}(S_{k}^{z})^{2},
\end{equation}
where $\textbf{S}_{i}$ represents a classical Heisenberg spin vector with unit magnitude, and
the former three sums are taken over the nearest-neighbor sites whereas remaining terms are taken
over all the lattice sites. In practice, spin structure of interface in such systems is experimentally inaccessible,
and it is quite a challenge to determine the magnetic nature near interface. Therefore, in order
to theoretically investigate the ferrimagnetic properties of the system, we consider the
following Hamiltonian parameters: Spin-spin interactions between core and shell spins are
taken as ferromagnetic $(J_{C,S}>0)$ whereas interactions at the core-shell
interface are of antiferromagnetic type $(J_{IF}<0)$. Due to the reduced
coordination number at the surface, FM exchange interactions between shell spins
are usually smaller than those  between core spins. Hence, we select $J_{S}=0.5 J_{C}$
with $|J_{IF}|\leq2.0J_{C}$. Easy axis magnetization of the particle is assumed to be
along the $z$ direction by assigning nonzero values for the $z$ components of the uniaxial anisotropy
constants $K_{C}$ and $K_{S}$ for the particle core and shell, respectively.
In order to emphasize the surface effects, we set $K_{C}=0.1J_{C}=0.1K_{S}$ \cite{kaneyoshi9}.

A variety of values for the total radius has been considered as $R=10.0,15.0,20.0,25.0$
and $30.0$ throughout the simulations, corresponding to the particles
with $N_{T}=4169,14147,33401,65267$ and $113081$ spins, respectively. However,
in most cases, the shell thickness has been varied at the expense of
the core which mimics the case of the production process for the surface
chemically modified nanoparticles \cite{Nogues}. Our simulations are
based on an improved Metropolis algorithm for classical Heisenberg spins
\cite{nowak,marsaglia,Evans3}, and we apply free boundary conditions in all directions.

Following physical quantities have been calculated in the simulation process:
\begin{itemize}
\item Thermal and configurational averages of the instantaneous magnetizations,
\begin{eqnarray}\label{eq2}
\nonumber
M_{C}^{\alpha}&=&\frac{1}{N_{C}}\left\langle\sum_{i=1}^{N_{C}}S_{i}^{\alpha}\right\rangle,\\
\nonumber
M_{S}^{\alpha}&=&\frac{1}{N_{S}}\left\langle\sum_{i=1}^{N_{S}}S_{i}^{\alpha}\right\rangle, \\
\nonumber
M_{IF}^{\alpha}&=&\frac{1}{N_{IF}}\left\langle\sum_{i=1}^{N_{IF}}S_{i}^{\alpha}\right\rangle,\\
M_{T}^{\alpha}&=&\frac{1}{N_{T}}\left\langle\sum_{i=1}^{N_{T}}S_{i}^{\alpha}\right\rangle \quad
\mathrm{with} \quad \alpha=x,y,z.
\end{eqnarray}
where $M_{C}^{\alpha}$, $M_{S}^{\alpha}$, $M_{IF}^{\alpha}$, $M_{T}^{\alpha}$
denote the average magnetizations corresponding to core, shell, interface parts,
and total system, respectively with corresponding number of lattice
sites $N_{C}$, $N_{S}$, $N_{IF}$ and $N_{T}$ which depend on the shell
thickness and the total radius of the particle. Hence, using Eq. (\ref{eq2}), the
magnitude of the magnetization vector can be computed via
\begin{equation}\label{eq3}
M_{\delta}=\sqrt{(M_{\delta}^{x})^2+(M_{\delta}^{y})^2+(M_{\delta}^{z})^2}, \quad \delta=C,S,IF \ \mathrm{or} \ T
\end{equation}
for any distinct region of the particle.
\item Internal energy and specific heat per lattice site can also be calculated using
\begin{equation}\label{eq4}
E_{\mathrm{total}}=-\langle \mathcal{H}\rangle/ N_{T}, \quad C=\frac{dE_{\mathrm{total}}}{dT}.
\end{equation}
\end{itemize}
In order to calculate the temperature dependence of magnetization and specific
heat curves, we have performed simulations with $5\times10^{4}$ steps at each temperature.
In order to allow the system to reach a stationary state, we have discarded the first $25\%$ steps
for thermalization. In the calculations, we set $k_{B}=1$ for simplicity.

\section{Results and discussion}\label{results}
\begin{figure*}[!sh]
\center
\subfigure[\hspace{0cm}] {\includegraphics[width=6cm]{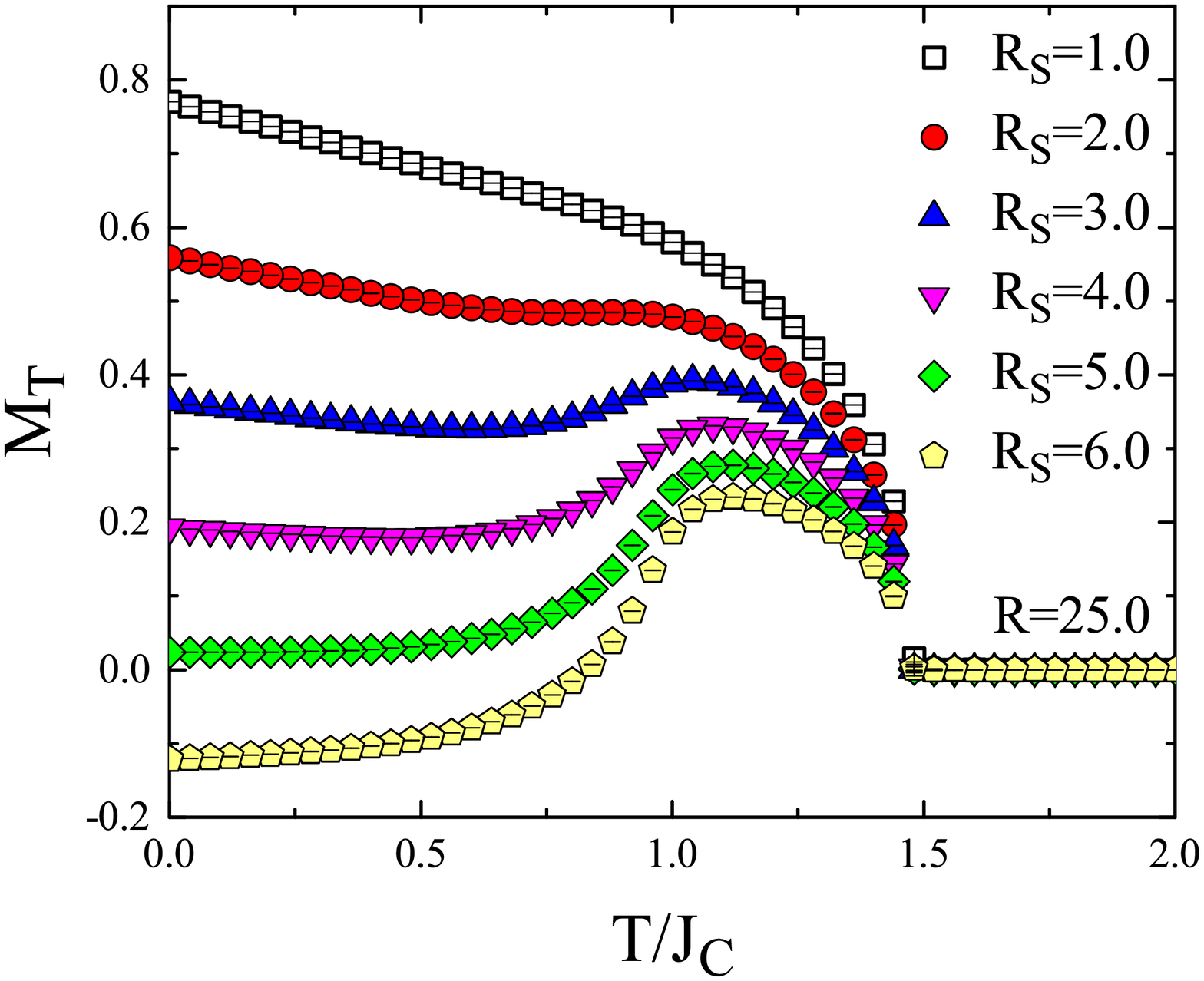}}
\subfigure[\hspace{0cm}] {\includegraphics[width=6cm]{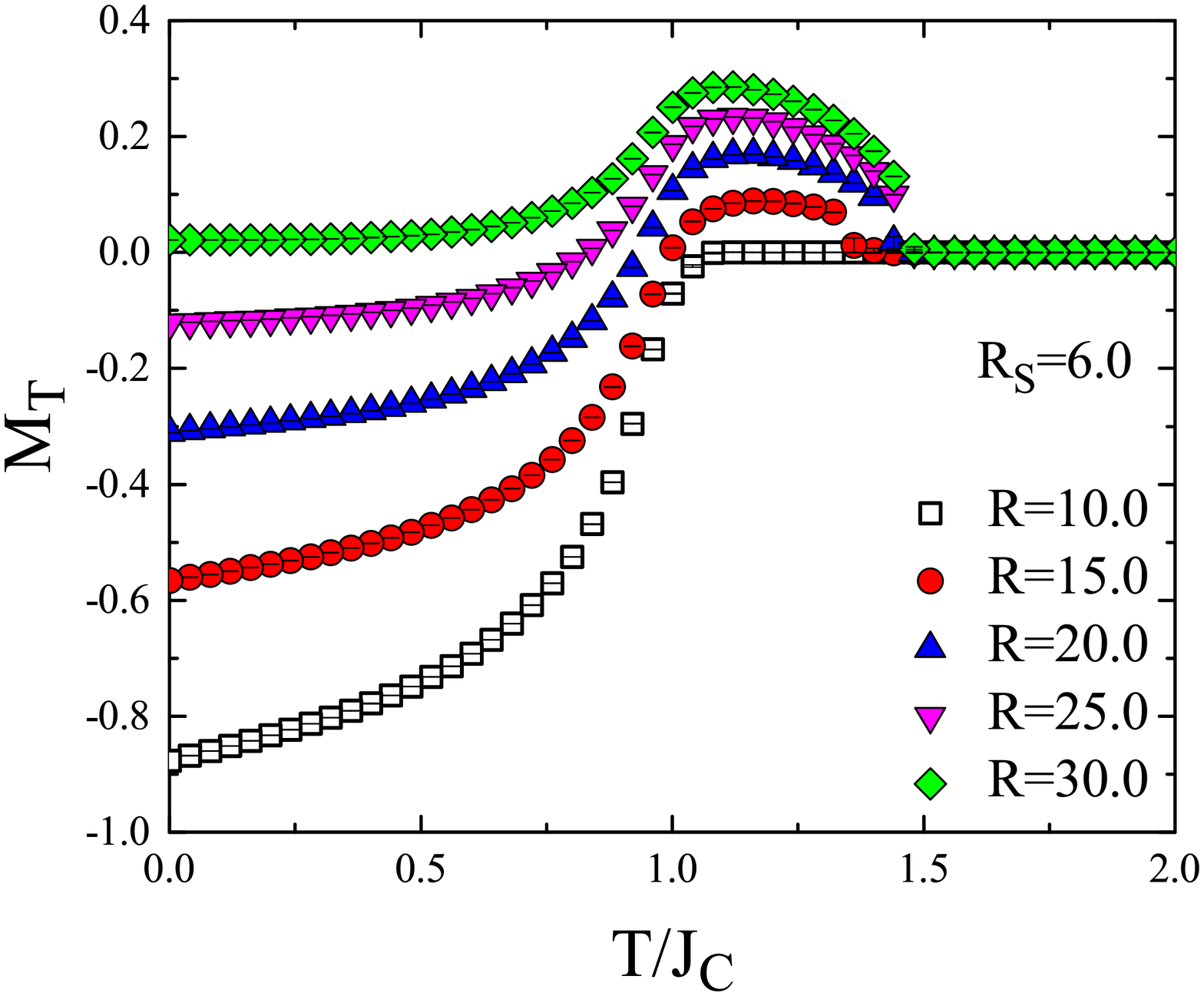}}\\
\subfigure[\hspace{0cm}] {\includegraphics[width=6cm]{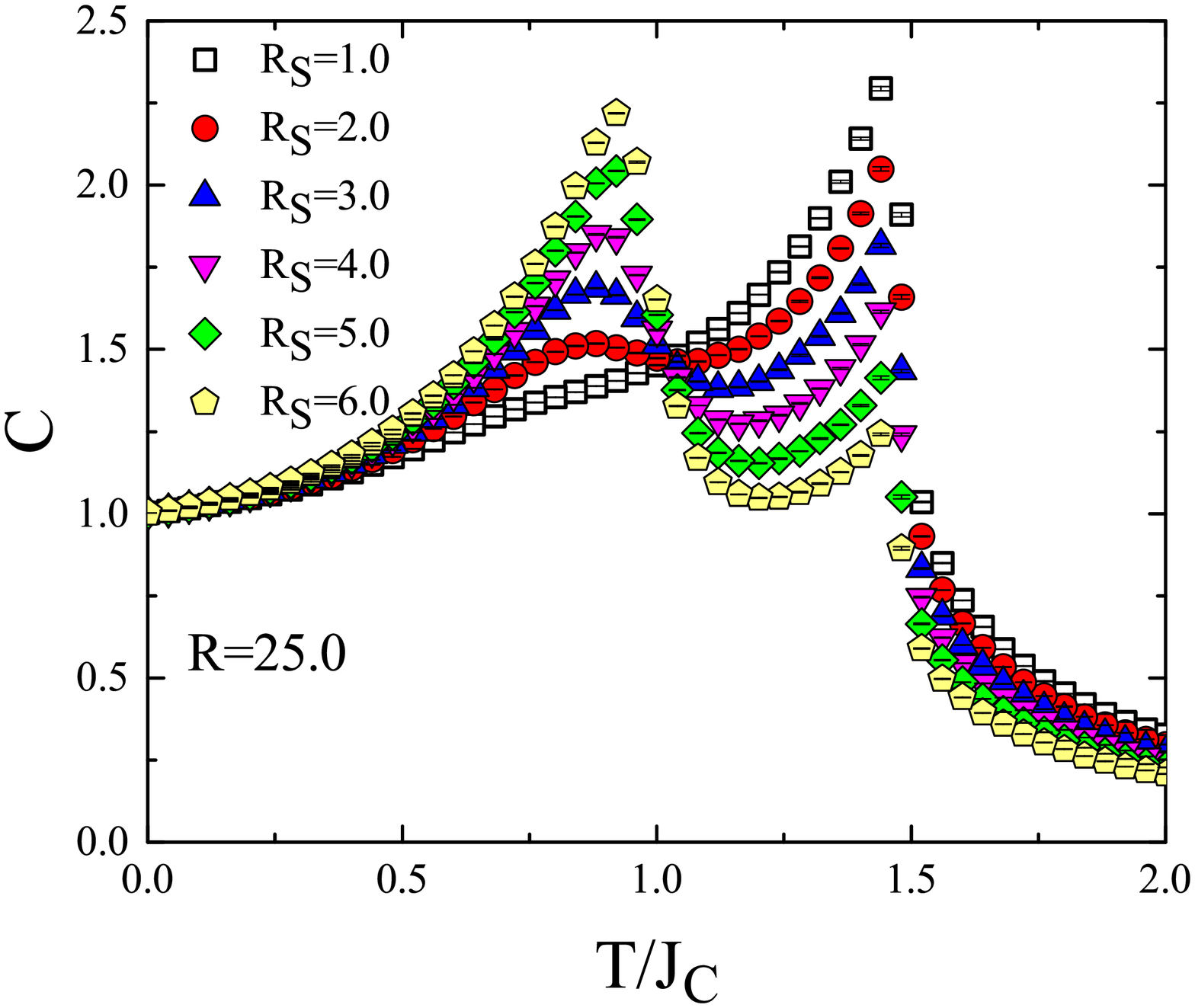}}
\subfigure[\hspace{0cm}] {\includegraphics[width=6cm]{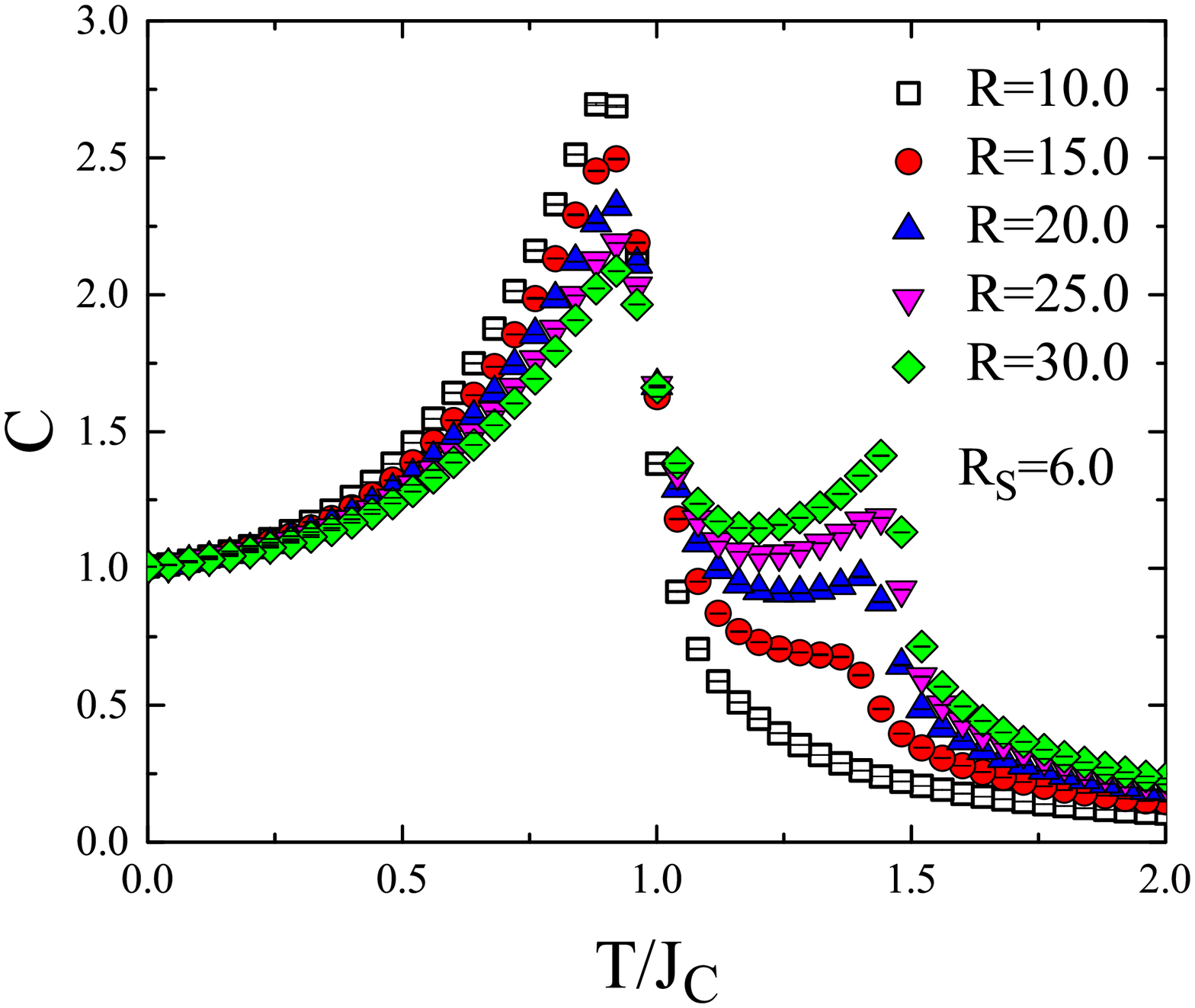}}\\
\caption{Temperature variation of total magnetization $M_{T}$ and specific heat $C$ corresponding
to simulated pure nanoparticle samples with some selected values of shell
thickness $R_{S}$ and total radius $R$. The results have been obtained for the
following system parameters: $J_{S}=0.5J_{C}$, $J_{IF}=-0.5J_{C}$, $K_{C}=0.1J_{C}$, $K_{S}=1.0J_{C}$.} \label{fig2}
\end{figure*}

In this section, let us discuss the results for the particles with disordered surface
and interfaces. However, before presenting these results, it would be beneficial
to take a glance at the magnetic properties of the pure system. The results for the
total magnetization $M_{T}$ and specific heat per spin $C$ as a function of
temperature, as well as the shell and overall particle sizes are shown in Fig. \ref{fig2}.
These results have been obtained after averaging over 10 different sample
realizations at each temperature. The error bars obtained via Jackknife method \cite{newman}  
are smaller than the data symbols. As shown in this
figure, nanoparticles with different structural parameters may exhibit different
magnetization profiles. Namely, according to the N\'{e}el classification scheme, thermal
dependence of magnetization curves can be classified in several categories. For instance,
as shown in Fig. \ref{fig2}a, the magnetization curves exhibit $Q,P,L,N$ type behaviors with
increasing shell thickness in successive order whereas for fixed shell thickness,
such as $R_{S}=6.0$, we obtain $Q,N,L,P$ type of behaviors, respectively for increasing total radius $R$,
(c.f. see Fig \ref{fig2}c). The existence of $N$ type magnetization profiles is
important in technological applications \cite{miyazaki}.

From Figs. \ref{fig2}c and \ref{fig2}d, it can be deduced that the specific heat
curves may exhibit two successive maximum points. Those observed at low temperature
region originates due to the thermal fluctuations in the shell, however the high
temperature peak is a consequence of the thermal variation of the core magnetization.
The magnitude of the both peaks are size dependent: core (shell) peak with a cusp-like
divergent behavior becomes a broadened hump as the core (shell) magnetization loses its
dominance with varying system size. We also note that the temperature value corresponding
to the peak located at high temperature region is the sole transition
temperature \cite{note2} since, the system becomes completely paramagnetic above this
temperature.

After presenting this general overview of the pure case, let us discuss the effects of
disorder due to the surface and interfacial impurities, as well as roughened interfaces.
Due to its technological importance, we will focus our attention on the variation of compensation
behavior \cite{note1} of the system in the presence of above mentioned disorder effects.

\subsection{Nanoparticles with Imperfect Shell Layers}\label{subsec1}

Nanoparticle system in the presence of surface disorder can be simulated by considering
randomly distributed non-magnetic sites on the surface and in the interior regions of
the shell layer (except the innermost shell sites which are located at the core-shell
interface region). In this case, the modified Hamiltonian defining the system reads

\begin{equation}\label{eq5}
\mathcal{H}=-J_{C}\sum_{<ij>}\textbf{S}_{i}.\textbf{S}_{j}-J_{S}\sum_{<kl>}\textbf{S}_{k}.\textbf{S}_{l}\zeta_{k}
\zeta_{l}-J_{IF}\sum_{<jk>}\textbf{S}_{j}.\textbf{S}_{k} 
-K_{C}\sum_{i}(S_{i}^{z})^{2}-K_{S}\sum_{k}\zeta_{k}(S_{k}^{z})^{2},
\end{equation}

\noindent where $\zeta_{k}$ is the site occupancy parameter which takes values zero or unity,
depending on whether the lattice site $k$ exhibits nonmagnetic or magnetic character, respectively.
In this manner, the total amount of disorder is controlled with a parameter $p$, such
that $p=1.0$ means all the lattice sites in the particle shell are magnetic whereas $p=0.0$ means that
no magnetic lattice points reside in the outer shell, and in this case, the only
contribution for the shell magnetization comes from the innermost shell spins (i.e. from the interfacial shell spins).

\begin{figure}
\includegraphics[width=6.0cm]{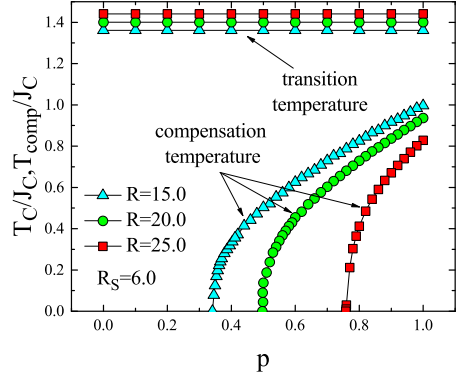}\\
\caption{The phase diagrams in a $(p-T_{C}/J_{C},T_{comp}/J_{C})$ plane for
the simulated particles with $R_{S}=6.0$, $R=15.0,20.0,25.0$. The other system parameters
are fixed as: $J_{S}=0.5J_{C}$, $J_{IF}=-0.5J_{C}$, $K_{C}=0.1J_{C}$, $K_{S}=1.0J_{C}$.}\label{fig3}
\end{figure}

Taking the above mentioned model as a basis, we plot the phase diagrams in a ($T_{C}/J_{C},T_{comp}/J_{C}$)
versus $p$ plane in Fig. \ref{fig3} for three different total radius values such
as $R=15.0$, $20.0$ and $25.0$ with a fixed shell thickness $R_{S}=6.0$. The horizontal data
in Fig. \ref{fig3} denotes the transition temperature of the system which is not
affected from the dilution of the shell region by non magnetic impurities.
As shown in Fig. \ref{fig2}b, this system exhibits a compensation point below the transition
temperature in the absence of disorder. By examining Fig. \ref{fig3}, one can see that
bigger particles exhibit lower compensation point $T_{comp}$ which decreases gradually
with increasing disorder, and vanishes at a critical $p_{c}$ value. Besides,
the compensation frontier in the phase diagram reaches to stronger disorder regime for smaller particles.

\begin{figure*}[!h]
\center
\subfigure[\hspace{0cm}] {\includegraphics[width=6cm]{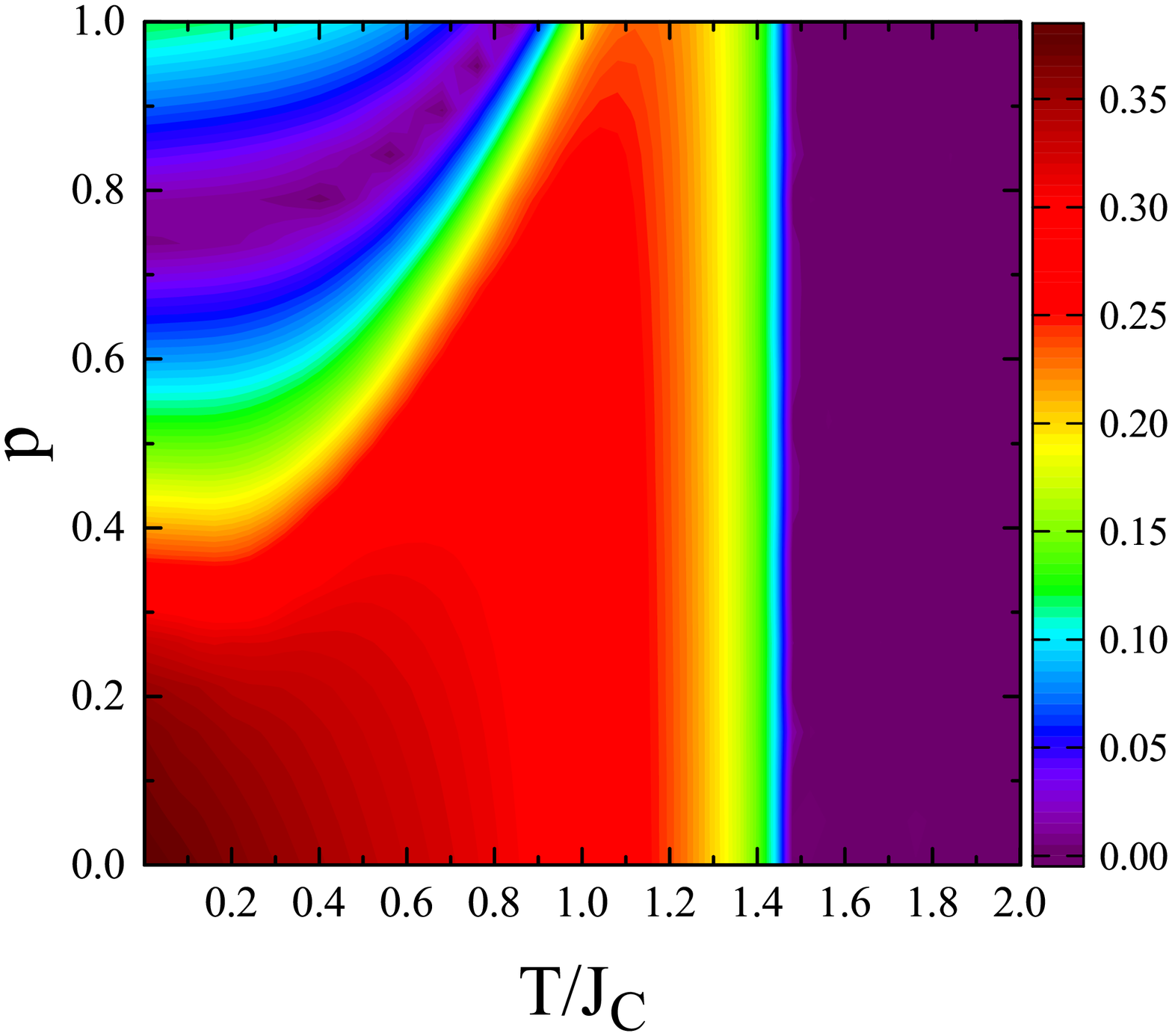}}
\subfigure[\hspace{0cm}] {\includegraphics[width=6cm]{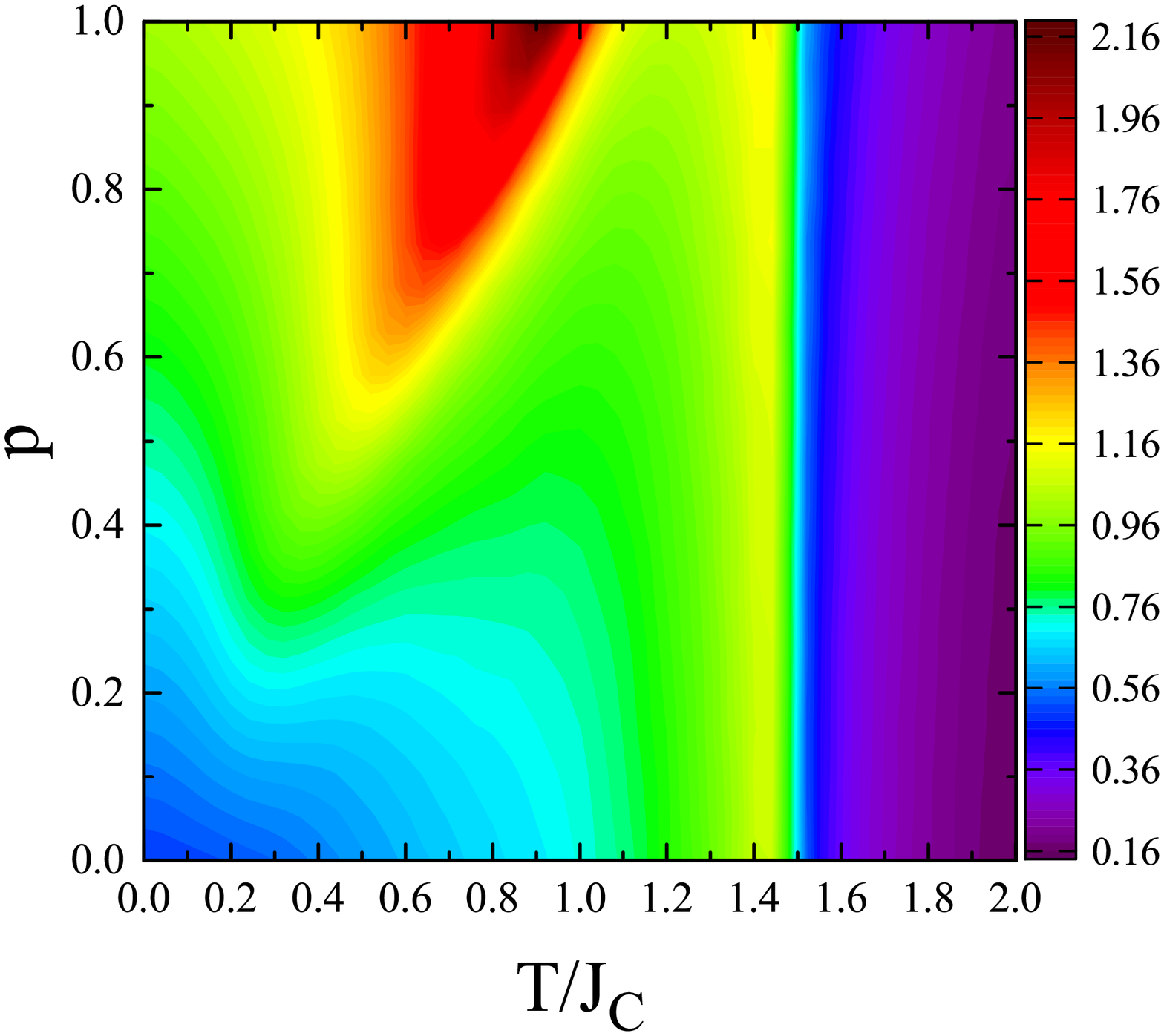}}\\
\caption{The color map contour plots of (a) the total magnetization $|M_{T}|$,
(b) specific heat $C$ for the system with $R=25.0$ and $R_{S}=6.0$ corresponding
to the phase diagram depicted in Fig. \ref{fig3}. The other fixed system parameters are as follows:
$J_{S}=0.5J_{C}$, $J_{IF}=-0.5J_{C}$, $K_{C}=0.1J_{C}$, $K_{S}=1.0J_{C}$.} \label{fig4}
\end{figure*}

The representative plots for color-maps of total magnetization $|M_{T}|$ and the
specific heat $C$ with $R=25.0$ and $R_{S}=6.0$ corresponding to the phase diagrams
depicted in Fig. \ref{fig3} are given in Fig. \ref{fig4}. According to Fig. \ref{fig4}a, the compensation
behavior originates in the low temperature and weak disorder region, since the particle size
is relatively big. At very low temperatures, the total magnetization can be
calculated from $|M_{T}|=|(N_{C}M_{C}+N_{S}M_{S})|/N_{T}$, hence it reaches its maximal value
when the shell magnetization is weak due to AFM exchange at the interface. In this context,
by taking the AFM structure of the interface exchange coupling into account,
we have $N_{C}=28671$, $N_{S}=36596$ for a perfectly pure particle $(p=1.0)$. Therefore,
it yields $|M_{T}|=0.121$, whereas for fully imperfect case $(p=0.0)$,
for $N_{C}=28671$ and $N_{S}=N_{S}^{IF}=3854$, we find  $|M_{T}|=0.38$ which agrees well
with the results shown in Fig. \ref{fig4}a. According to the N\'{e}el classification
scheme \cite{Neel}, $M_{T}$ exhibits $N$, $L$, $P$, and $Q$ type of behaviors with
increasing disorder. From Fig. \ref{fig4}b, it is clear that the specific heat
exhibits two successive peaks for $p=1.0$. However, as the amount of magnetic lattice sites
in the shell progressively decreases then the lower temperature peak tends to disappear.
Apart from these, both from Fig. \ref{fig4}a and \ref{fig4}b, one should notice that
the transition temperature of the system is located around $T_{C}\approx1.5J_{C}$,
and it is insensitive to the presence of shell disorder, since the major contribution
to the magnetism of the particle mainly comes from the core part.

\subsection{Nanoparticles with Diluted Interface}\label{subsec2}
\begin{figure*}[!h]
\center
\subfigure[\hspace{0cm}] {\includegraphics[width=5.0cm]{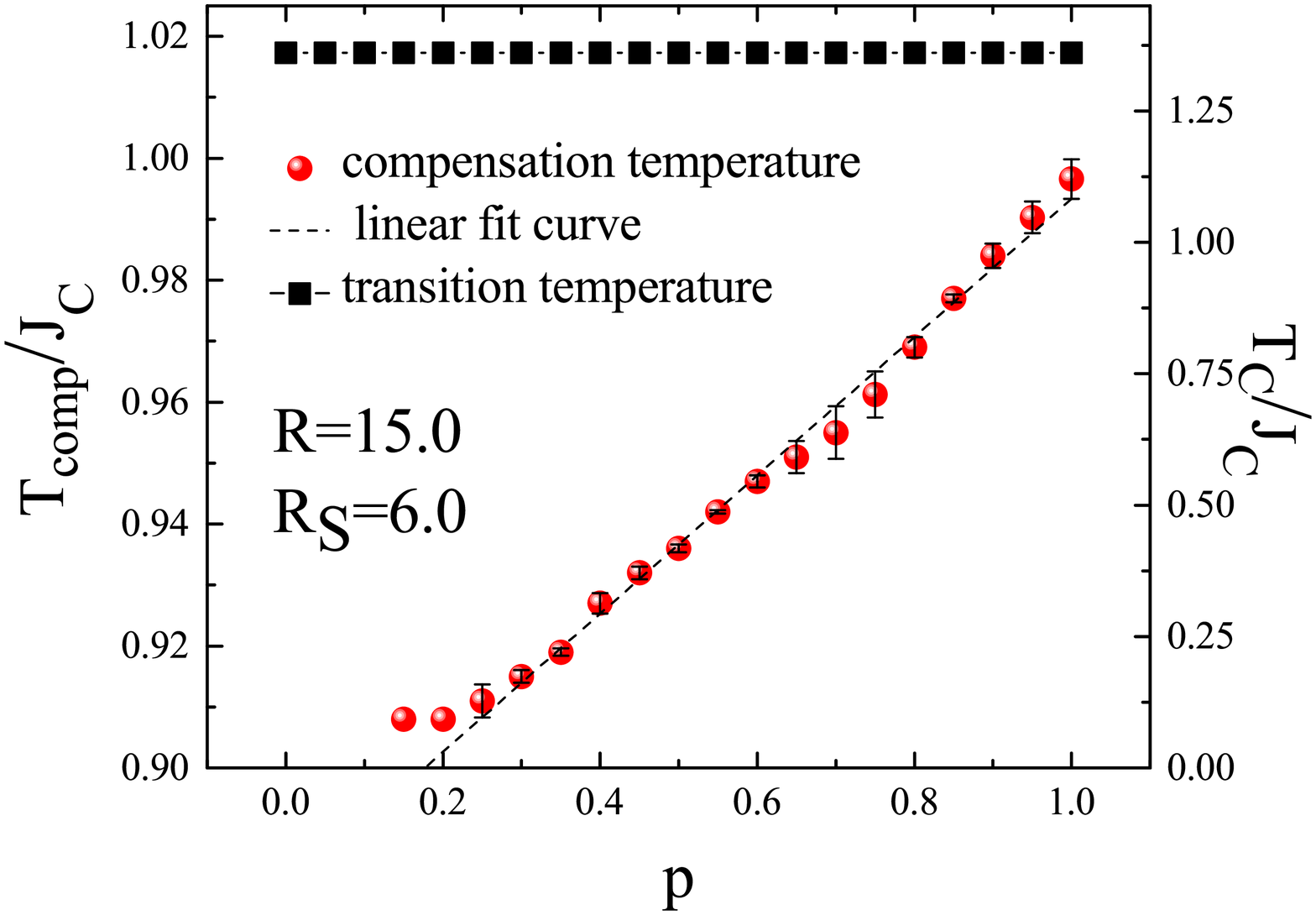}}
\subfigure[\hspace{0cm}] {\includegraphics[width=5.0cm]{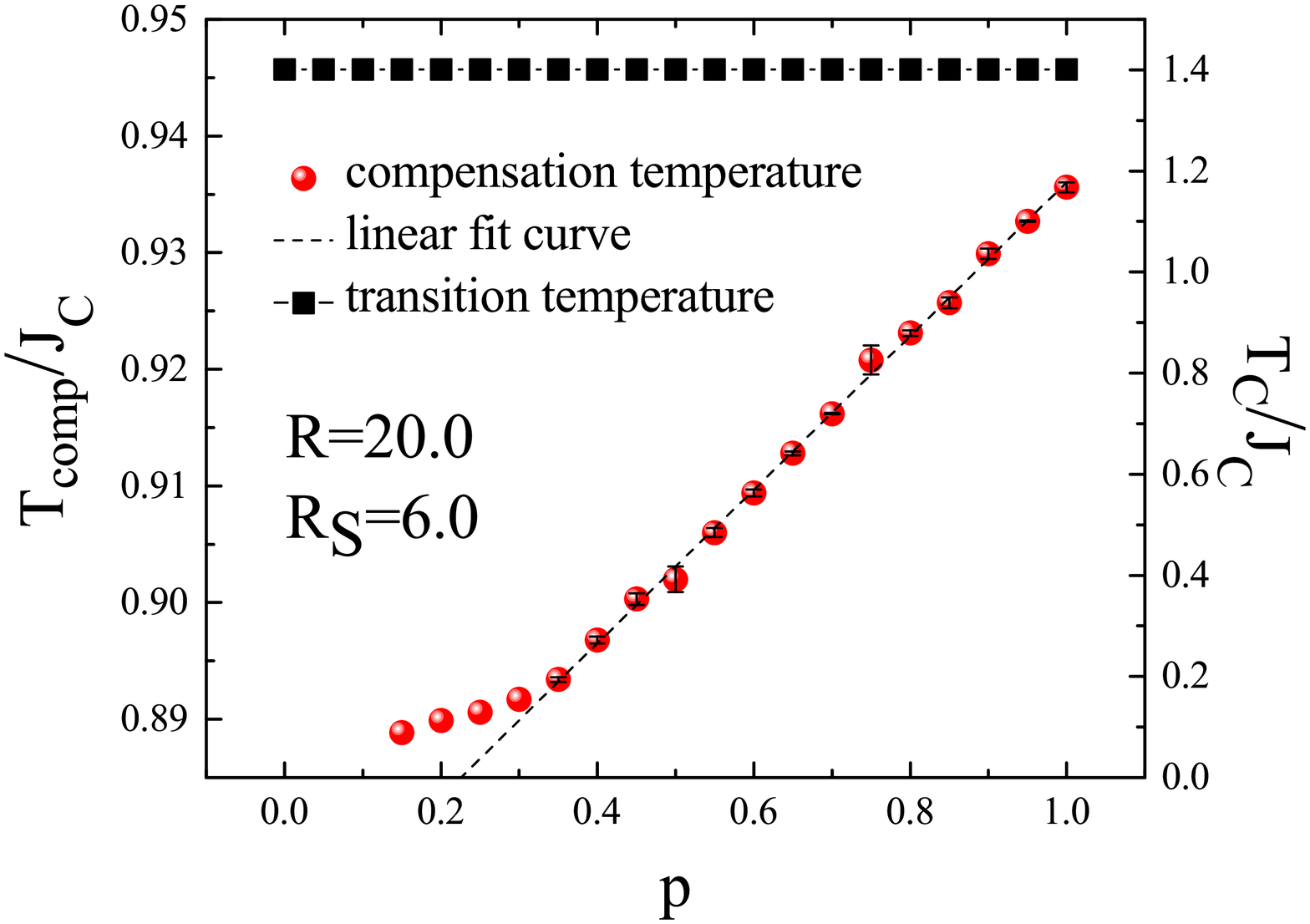}}
\subfigure[\hspace{0cm}] {\includegraphics[width=5.0cm]{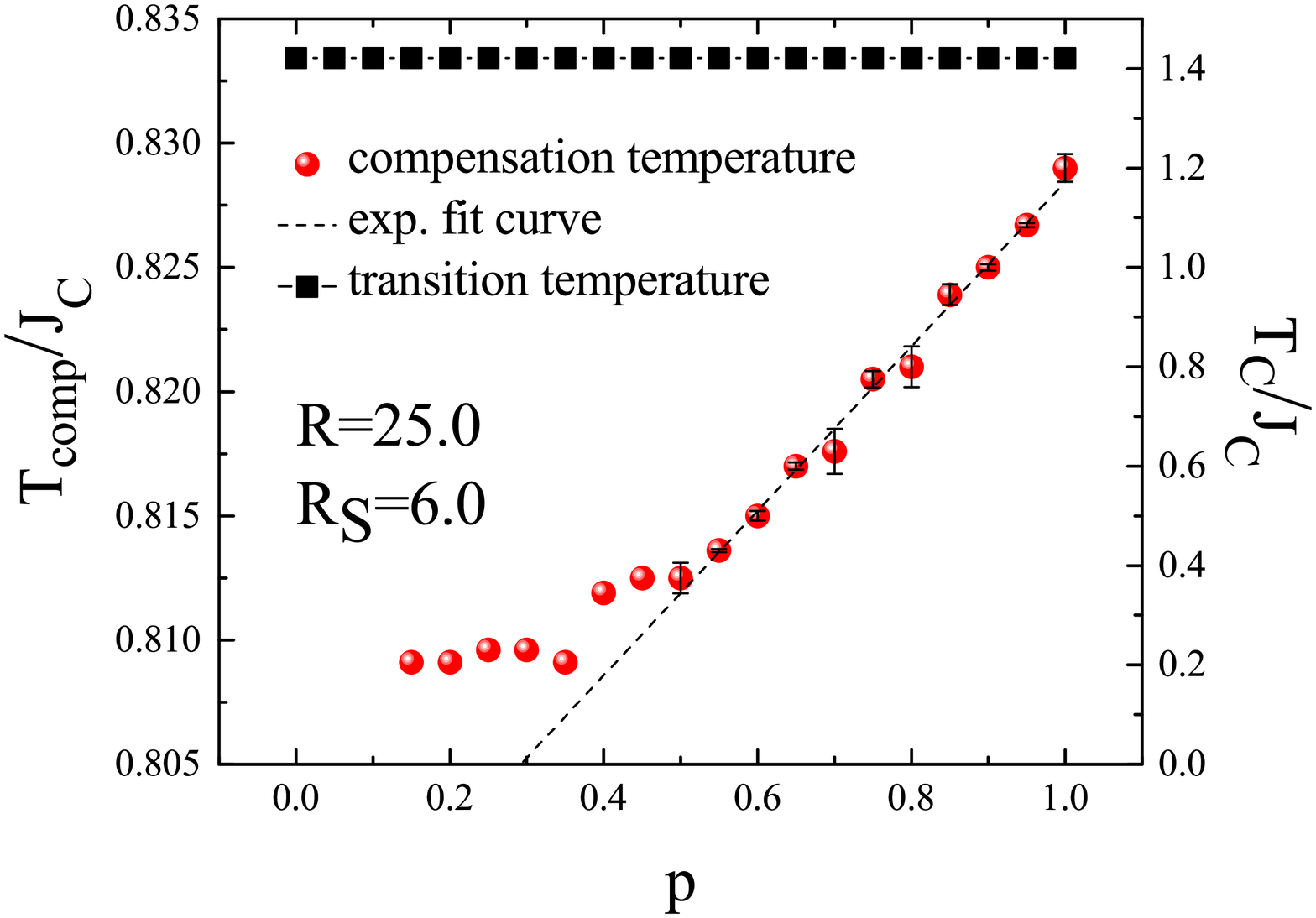}}\\
\caption{Phase diagrams in a $(p-T_{comp}/J_{C},T_{C}/J_{C})$ plane for simulated
particles with shell thickness $R_{S}=6.0$ with three different values of the total
radius (a) $R=15$, (b) $R=20$, (c) $R=25$. The other system parameters are fixed
as $J_{S}=0.5J_{C}$, $J_{IF}=-0.5J_{C}$, $K_{C}=0.1J_{C}$, $K_{S}=1.0J_{C}$. The
horizontal data symbols denote the transition temperature.} \label{fig5}
\end{figure*}

In this subsection, we discuss and present results for nanoparticles with disordered
interface sites. In this regard, one should modify the Hamiltonian given in Eq. (\ref{eq1}) as

\begin{equation}\label{eq6}
\mathcal{H}=-J_{C}\sum_{<ij>}\textbf{S}_{i}.\textbf{S}_{j}\zeta_{j}-J_{S}\sum_{<kl>}
\textbf{S}_{k}.\textbf{S}_{l}\zeta_{k}-J_{IF}\sum_{<jk>}\textbf{S}_{j}.\textbf{S}_{k}\zeta_{j}\zeta_{k} 
-K_{C}\sum_{i}\zeta_{i}(S_{i}^{z})^{2}-K_{S}\sum_{k}\zeta_{k}(S_{k}^{z})^{2},
\end{equation}

\noindent where $\zeta_{k}$ is the site occupancy parameter, as usual. $p$ is the ratio of magnetic sites
located at the core-shell interface with $p=1.0$ and $0.0$, respectively corresponding to
the cases with pure and maximally disordered situations. Fig. \ref{fig5} shows the variation of
compensation point $T_{comp}$ and transition temperature $T_{C}$ as functions of $p$ for
three different values of the total radius $R$ of the particle. This figure shows
that $T_{comp}$ always decreases, but the transition temperature does not change with
increasing interface disorder. There exists a region in the disorder spectrum
such as $1.0\geq p\geq p^{*}$ where $T_{comp}$ linearly decreases with decreasing $p$ such
as $T_{comp}\propto \alpha p$, and the dashed lines accompanying the simulation data
denote the linear fitting curves. The linear variation region becomes narrower for bigger particles.
From our data we find $\alpha=0.113$, $0.065$, and $0.033$ for respective radius values
$R=10.0$, $15.0$, and $25.0$. The statistical error due to fitting procedure is
in the order of $10^{-15}$ in all cases.

\begin{figure*}[!h]
\center
\subfigure[\hspace{0cm}] {\includegraphics[width=6cm]{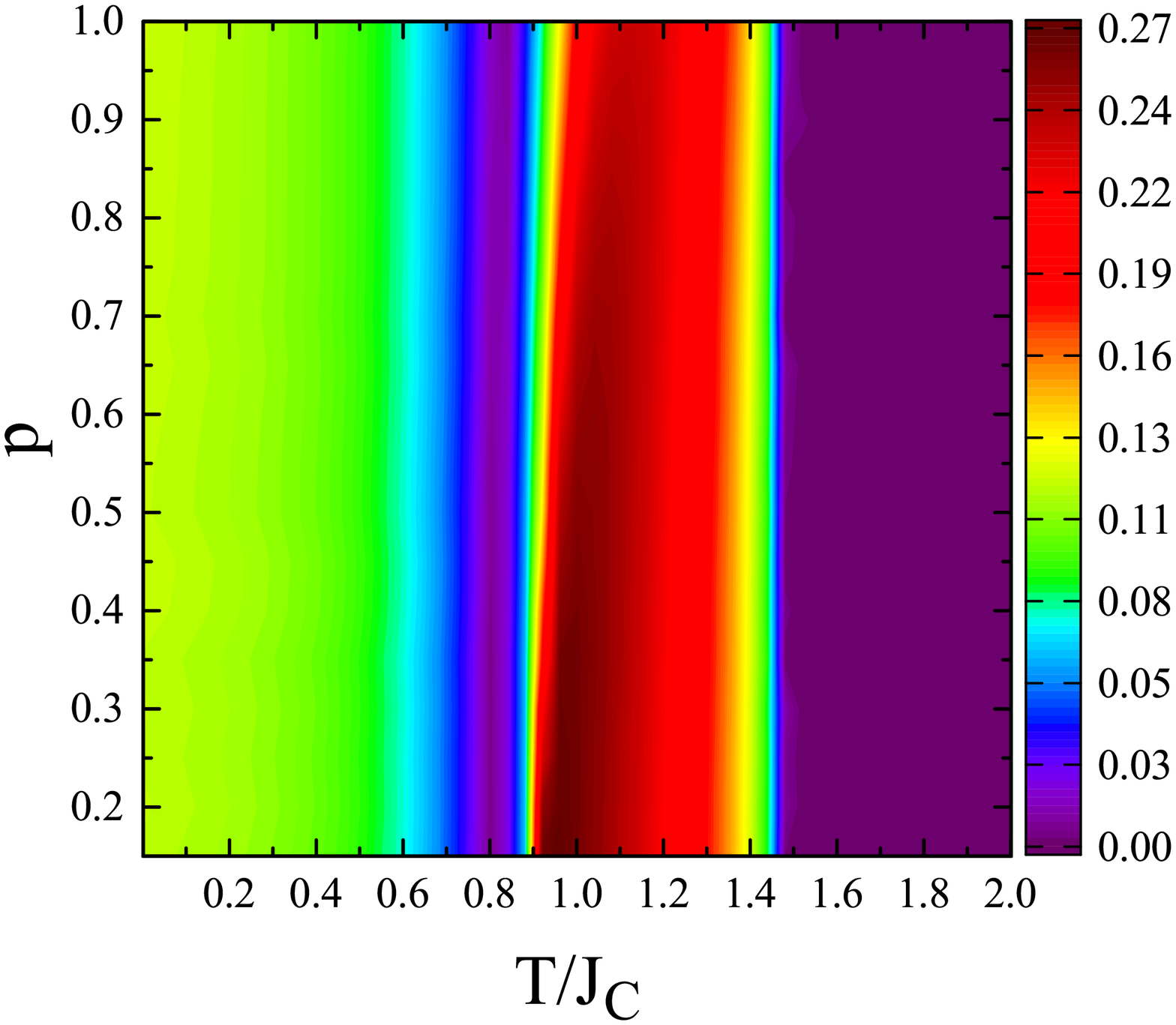}}
\subfigure[\hspace{0cm}] {\includegraphics[width=6cm]{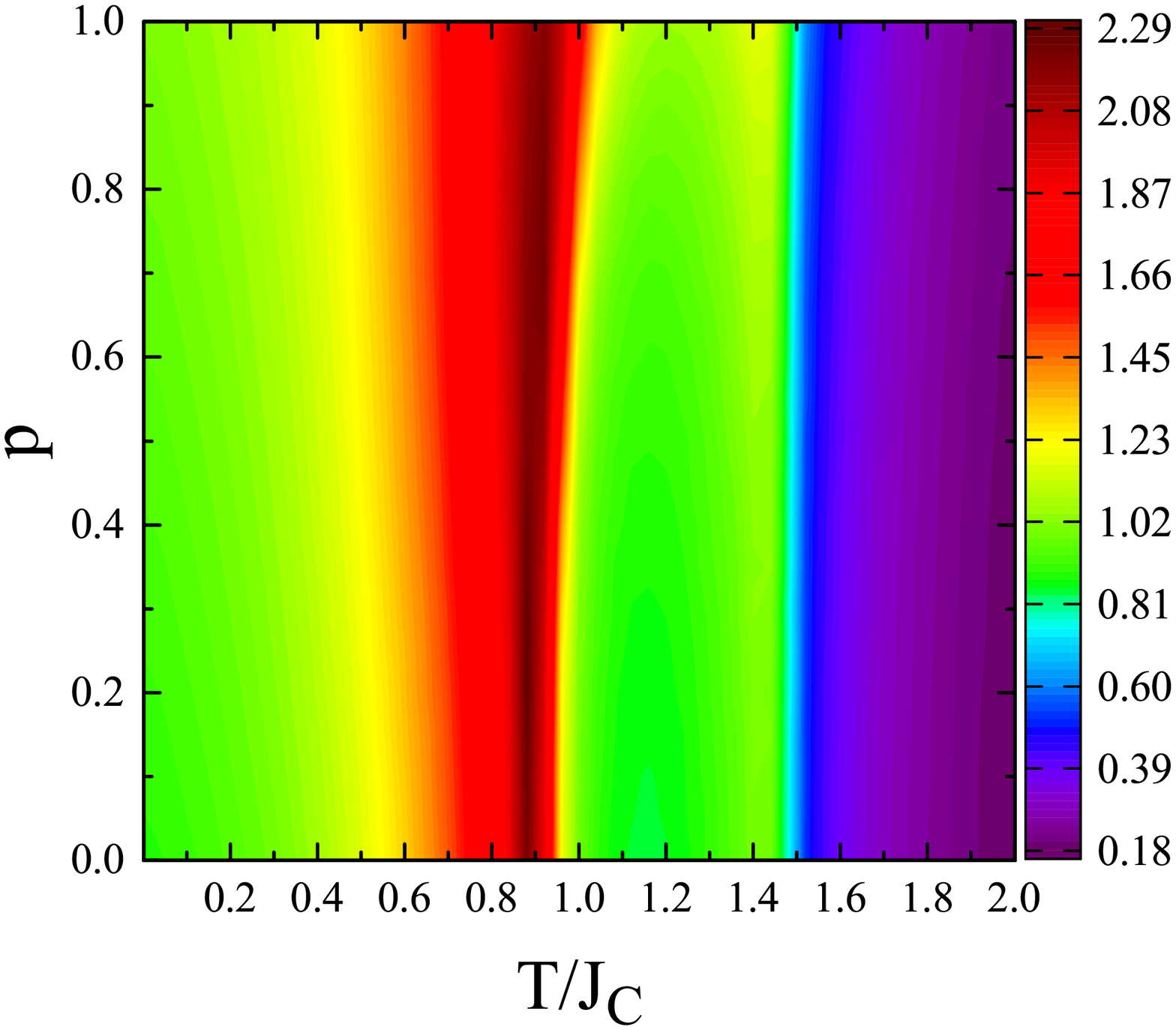}}\\
\caption{The color map contour plots of (a) the total magnetization $|M_{T}|$, (b)
specific heat $C$ for the system with $R=25.0$ and $R_{S}=6.0$ corresponding to
the phase diagram depicted in Fig. \ref{fig5}c. The other fixed system parameters are as follows:
$J_{S}=0.5J_{C}$, $J_{IF}=-0.5J_{C}$, $K_{C}=0.1J_{C}$, $K_{S}=1.0J_{C}$.} \label{fig6}
\end{figure*}

The representative contour plots regarding the magnetization $|M_{T}|$ and  specific heat $C$ as
functions of temperature and  parameter $p$ are shown in Fig. \ref{fig6}
for $R=25.0$ and $R_{S}=6.0$ corresponding to the phase diagrams displayed in
Fig. \ref{fig5}c. The variation of $T_{comp}$ as a function of $p$
around $T=0.8J_{C}$ is not clearly visible since the constant of proportionality $\alpha$ is very small.
However, the system always maintains its $N$ type magnetization with varying $p$.
At the ground state, when $p=1.0$, the total magnetization $|M_{T}|$ is given by $|M_{T}|=|N_{C}M_{C}+N_{S}M_{S}|/N_{T}=0.121$
as before, (see Fig. \ref{fig4}). On the other hand, in the maximally
disordered case, among the number of $N_{IF}=7448$ sites of the
interface region, a number of $N_{C}^{IF}=3594$ and $N_{S}^{IF}=3854$ sites belonging
to core and shell interface regions are nonmagnetic. Hence the total
magnetization reads $|M_{T}|=|[(28671-3594)-(36596-3854)]/65267|=0.117$ which is in accordance
with the results given in Fig. \ref{fig6}a. Apart from these, specific heat curves exhibit two
successive peaks due to the reasons described in our previous discussions.

\subsection{Nanoparticles with Roughened Interface}\label{subsec3}
As a final investigation in this work, we consider magnetic nanoparticles with
roughened interfaces in the following discussions.  In order to simulate particles
with rough interfaces we use a method very similar to those utilized in previous
works \cite{Evans1, Dimitriadis}. Namely, for a given shell thickness and particle radius, we
determine the radius $R_{C}$ of the ferromagnetic core. Then we generate a local  radius $R_{C}^{i}$
value for each interface lattice site $i$. This local radius is drawn from a Gaussian distribution  which has
variance $\sigma$, and it is centered at $R_{C}$. In this way, we sample number of $N_{IF}$ different  random
variables. If the local radius $R_{C}^{i}$ of a particular interface lattice site $i$ satisfies
the inequality $R_{C}+\delta\leq R_{C}^{i} \leq R_{C}-\delta$ then this lattice site remains unaltered.
In other words, if it belongs to the core interface, it remains in there throughout the generation
process for the particle shape (in our simulations, the parameter $\delta$ corresponds to $10\%$ of the unitary
lattice spacing). On the other hand, if $R_{C}^{i}$ does not satisfy the above inequality, then we generate
a random number $r$ within the interval $[0,1)$. If $r>0.5$ then the $i^{th}$ lattice site belongs to
the core interface, else it is assigned to the shell interface. Using this process, amount of roughness
is controlled by varying the Gaussian distribution width $\sigma$, and we successfully create roughened
interfaces without producing isolated nonmagnetic sites \cite{Dimitriadis}.

\begin{table*}[h!]
\begin{center}
\begin{threeparttable}
\caption{Structural parameters of simulated nanoparticles with ideal interface structure. } \label{table1}
\renewcommand{\arraystretch}{1.0}
\begin{tabular}{cccccccccc}
\thickhline
Sample & $R$ \ & $R_{S}$ \  & $N_{C}$ \  & $N_{S}$ \ & $N_{int}$ \  & $N_{C}^{int}$ \  & $N_{S}^{int}$ \  & $N_{T}$    \\
\hline
 1 & 25.0 & 2.0 & 50883 & 14384 & 5354 & 5670  & 11024 & 65267  \\
 2 & 25.0 & 6.0 & 28671 & 36596 & 3594 & 3854  & 7448 & 65267  \\
 3 & 15.0 & 2.0 & 9171 & 4976 & 1674 & 1854  & 3528 & 14147 \\
 4 & 15.0 & 6.0 & 3071 & 11076 & 794& 918 & 1712 & 14147  \\
\thickhline \\
\end{tabular}
\end{threeparttable}
\end{center}
\end{table*}

\begin{figure*}[!h]
\center
\subfigure[\hspace{0cm}] {\includegraphics[width=6cm]{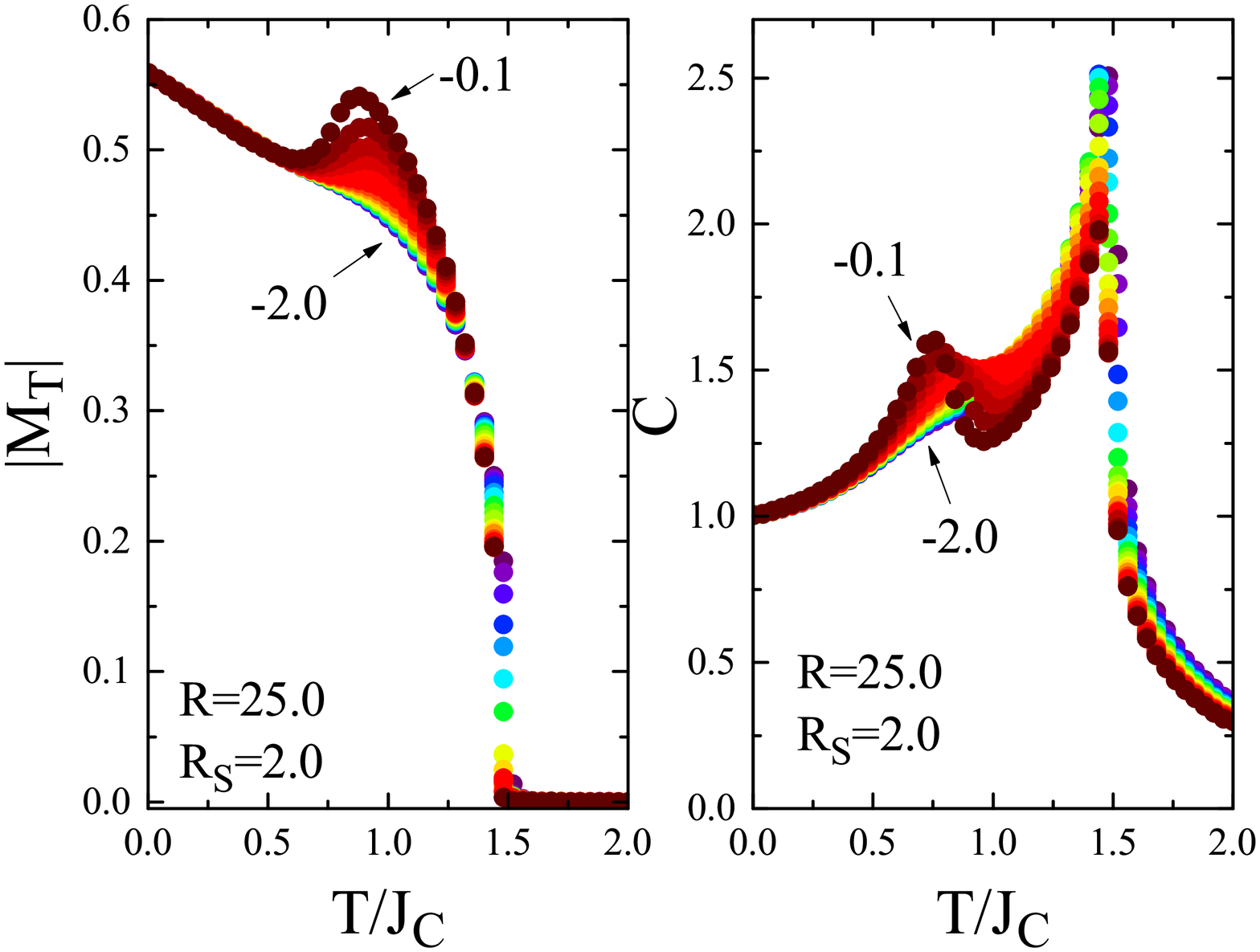}}
\subfigure[\hspace{0cm}] {\includegraphics[width=6cm]{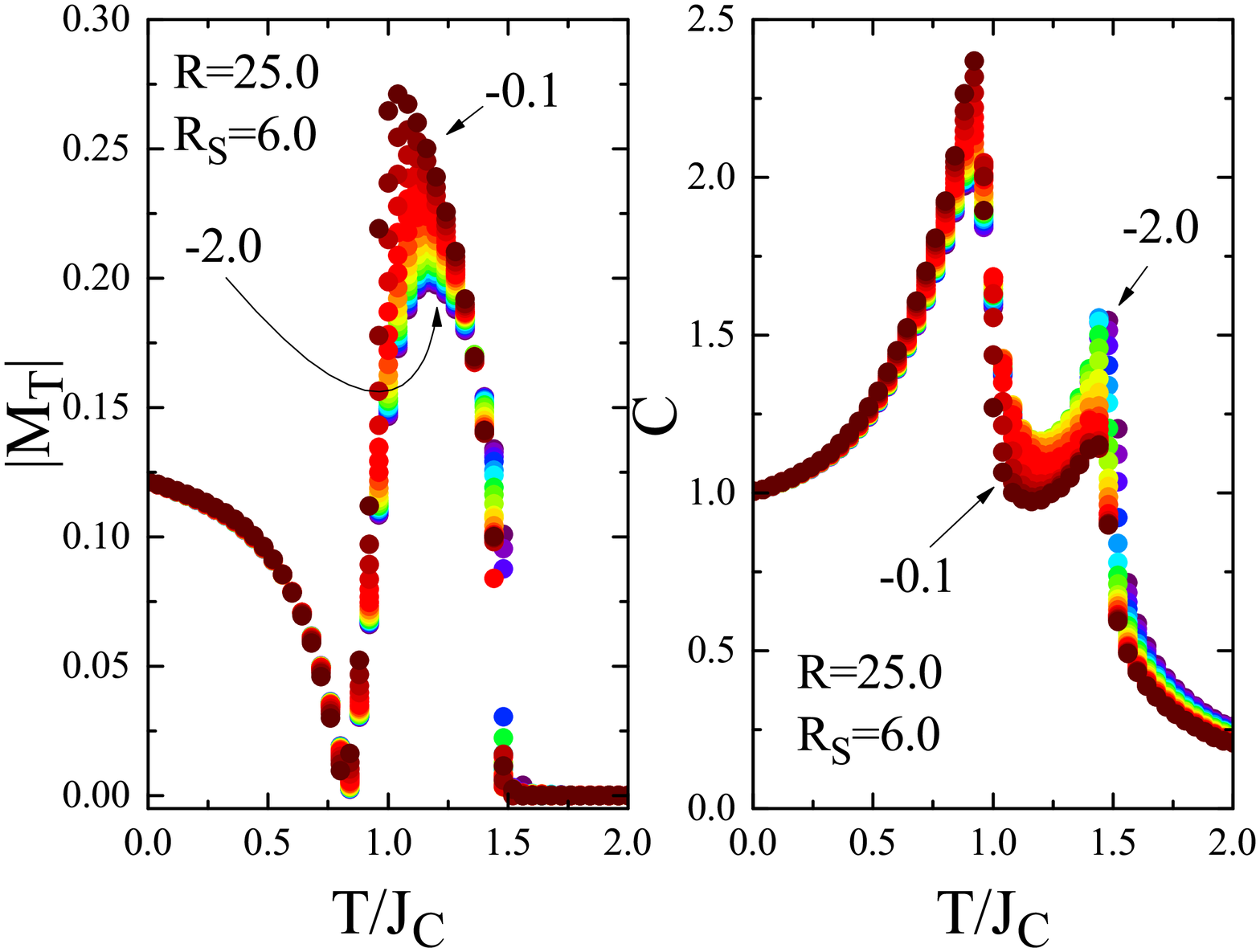}}\\
\subfigure[\hspace{0cm}] {\includegraphics[width=6cm]{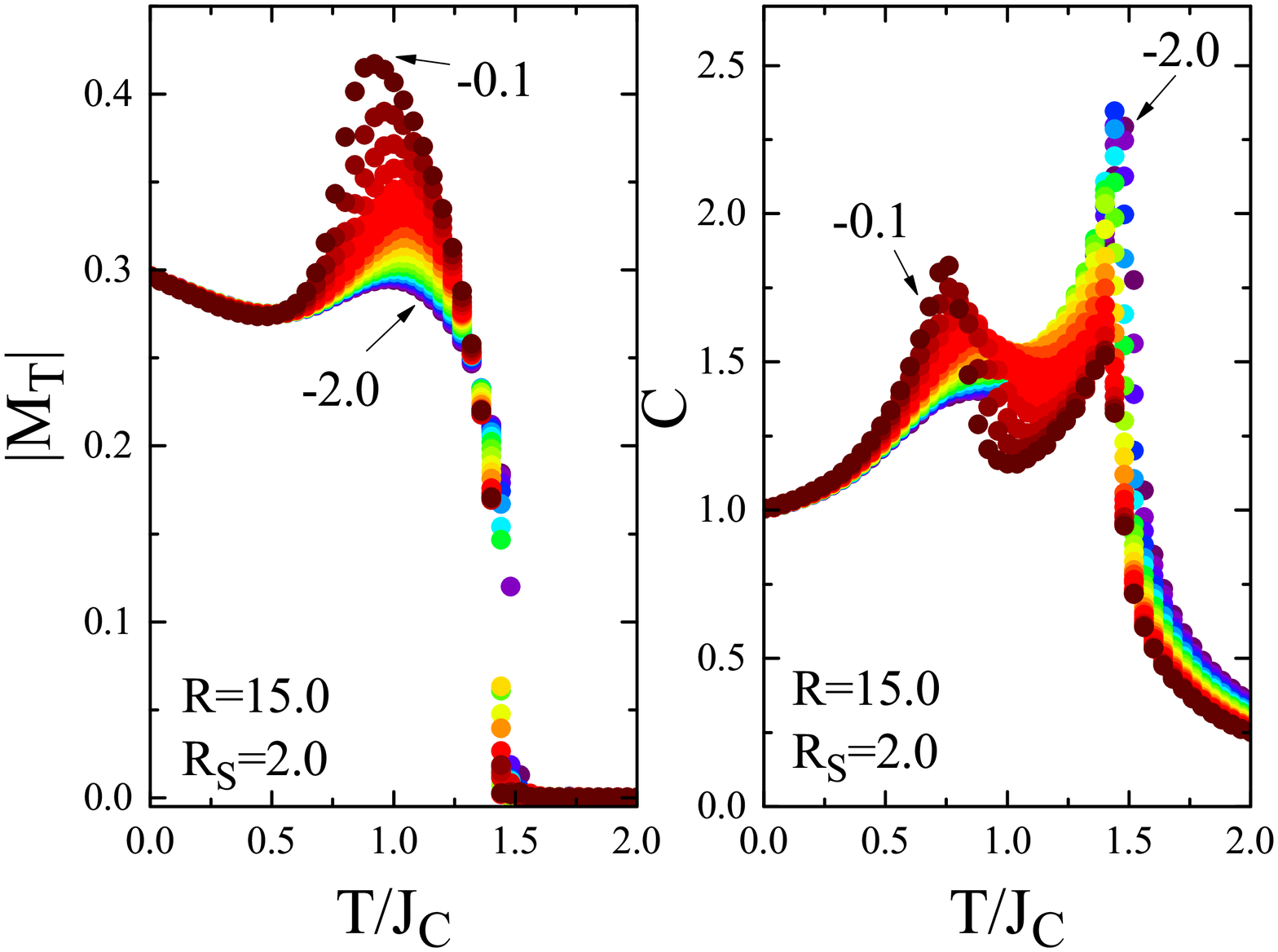}}
\subfigure[\hspace{0cm}] {\includegraphics[width=6cm]{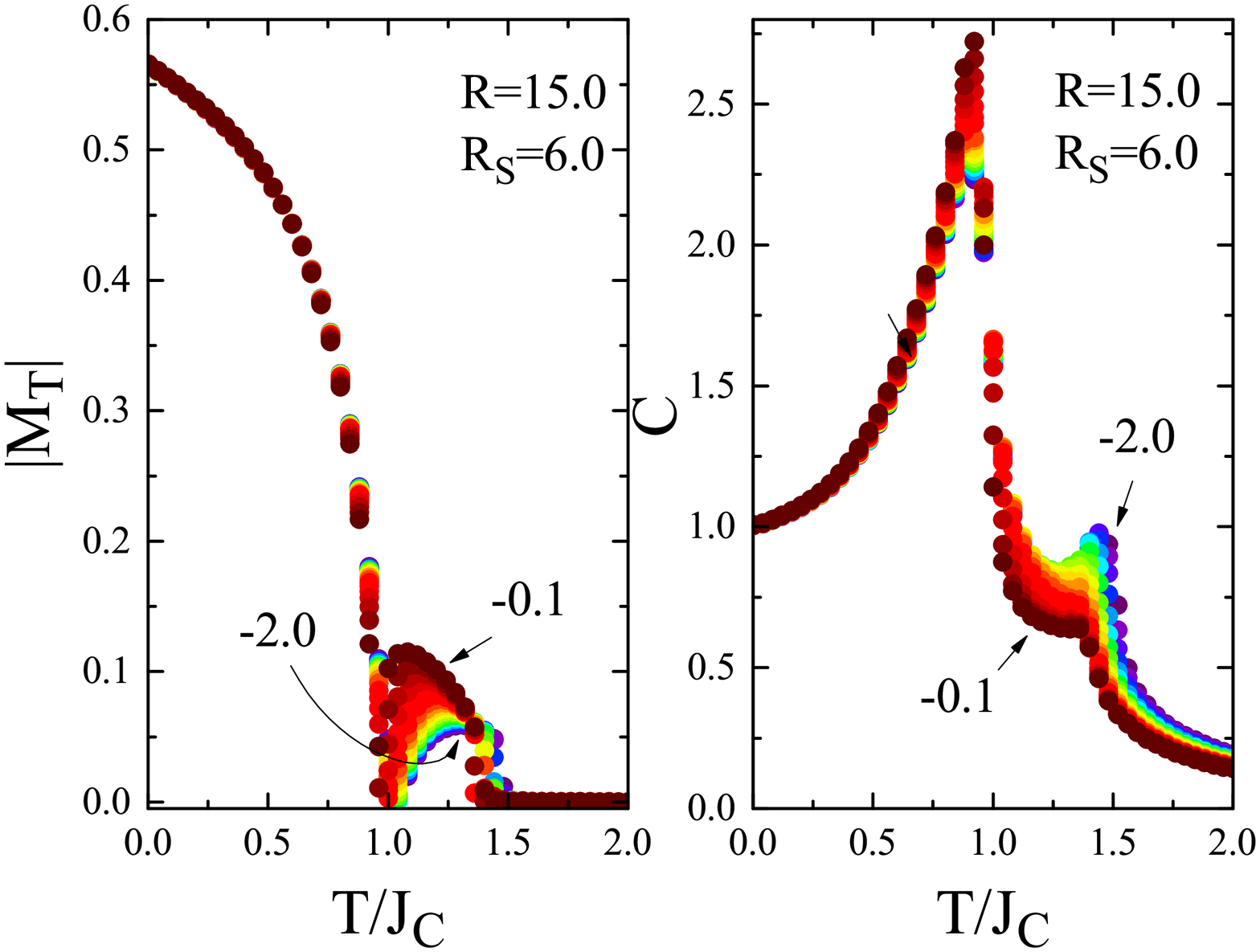}}\\
\caption{Temperature dependence of the total magnetization $|M_{T}|$ and
specific heat $C$ for a wide variety of interfacial exchange  coupling $-0.1\geq J_{IF}\geq -2.0$.
Four different samples have been prepared for simulation: (a) Sample 1
with $R=25.0$, $R_{S}=2.0$, (b) Sample 2 with $R=25.0$, $R_{S}=6.0$,
(c) Sample 3 with $R=15.0$, $R_{S}=2.0$, and (d) Sample 4 with $R=15.0$, $R_{S}=6.0$.
The other system parameters are fixed as $J_{S}=0.5J_{C}$, $K_{C}=0.1J_{C}$, $K_{S}=1.0J_{C}$.} \label{fig7}
\end{figure*}

\begin{figure*}[!h]
\center
\subfigure[\hspace{0cm}]{\includegraphics[width=4.cm]{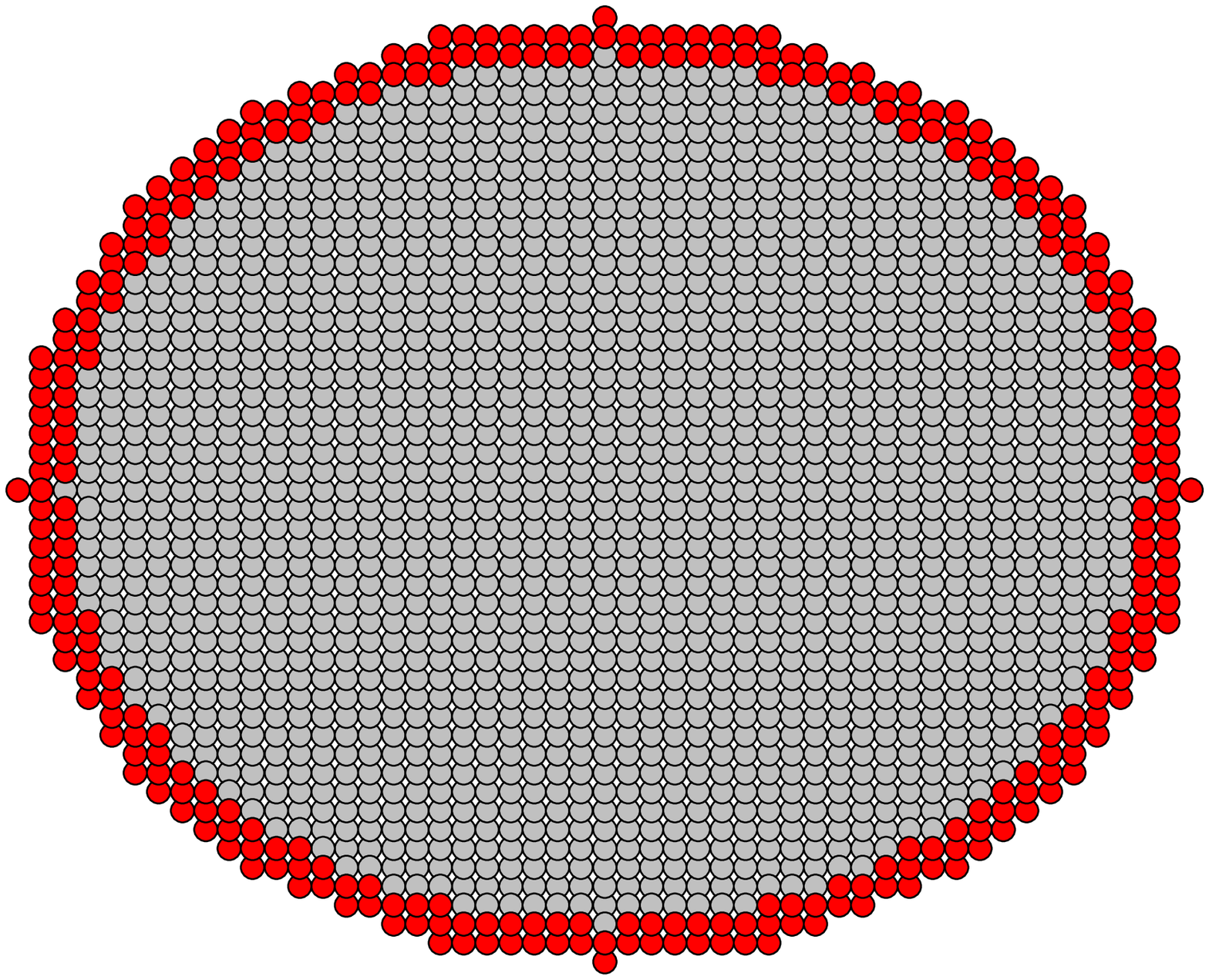}}
\subfigure[\hspace{0cm}]{\includegraphics[width=4.cm]{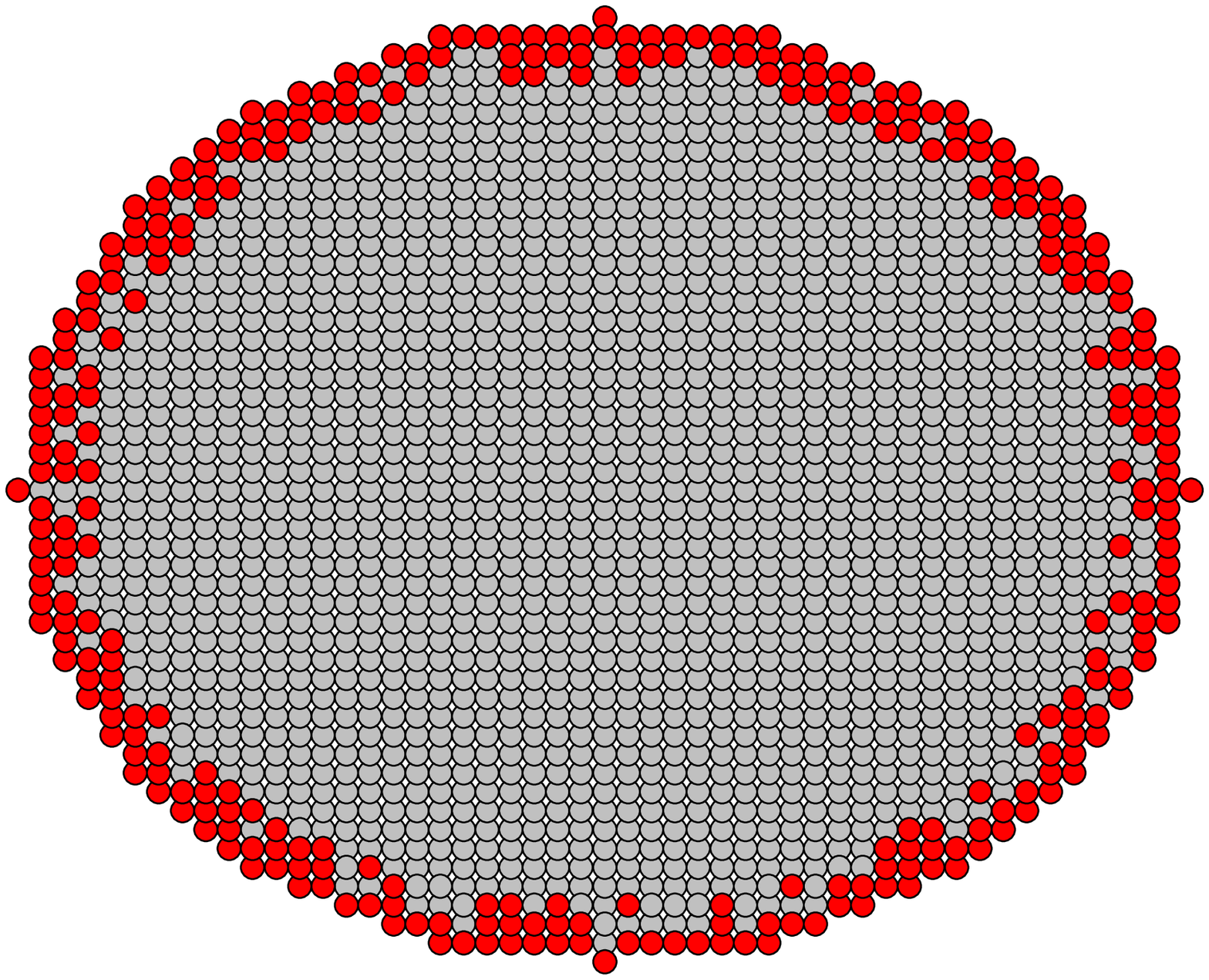}} \\
\subfigure[\hspace{0cm}]{\includegraphics[width=4.cm]{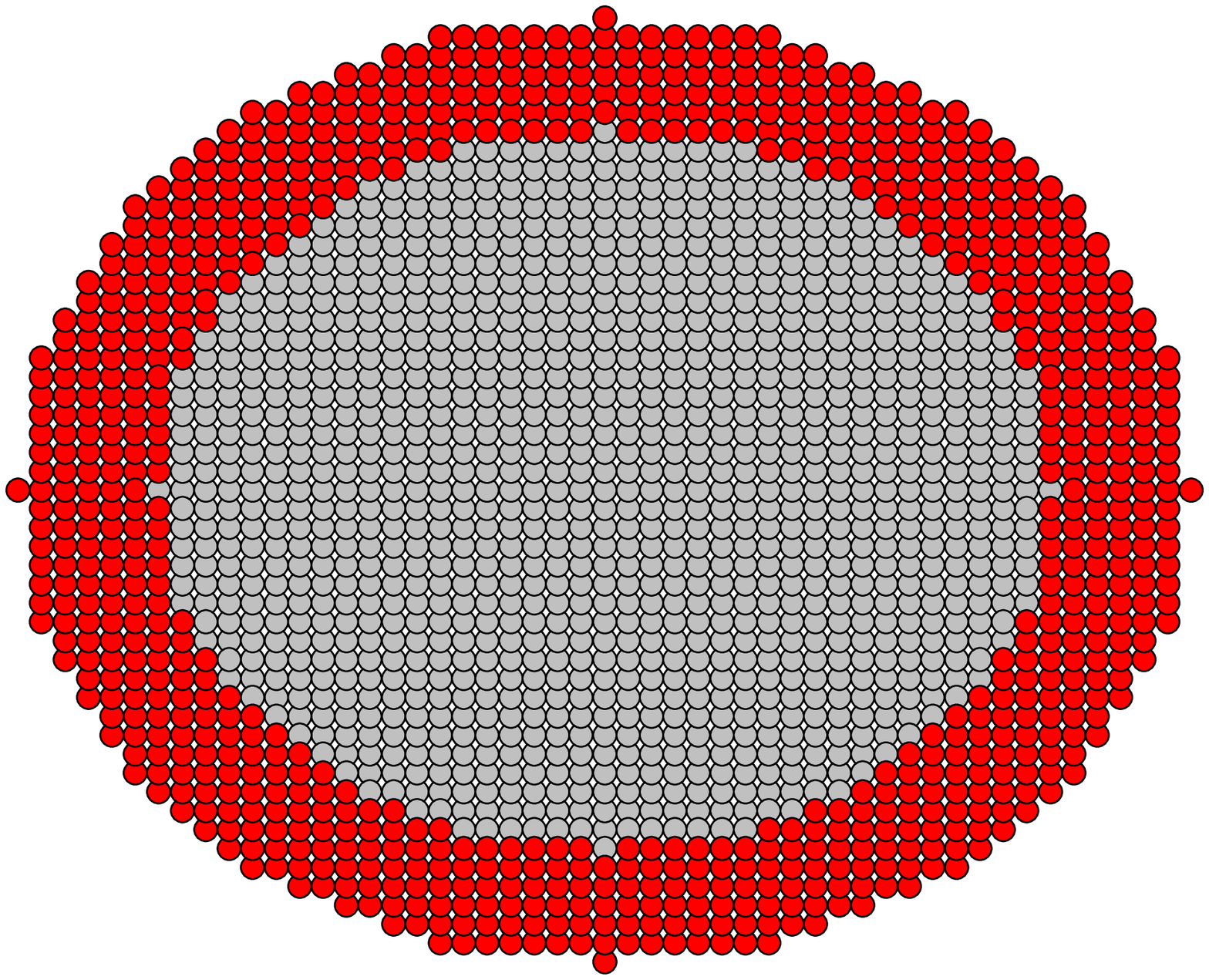}}
\subfigure[\hspace{0cm}]{\includegraphics[width=4.cm]{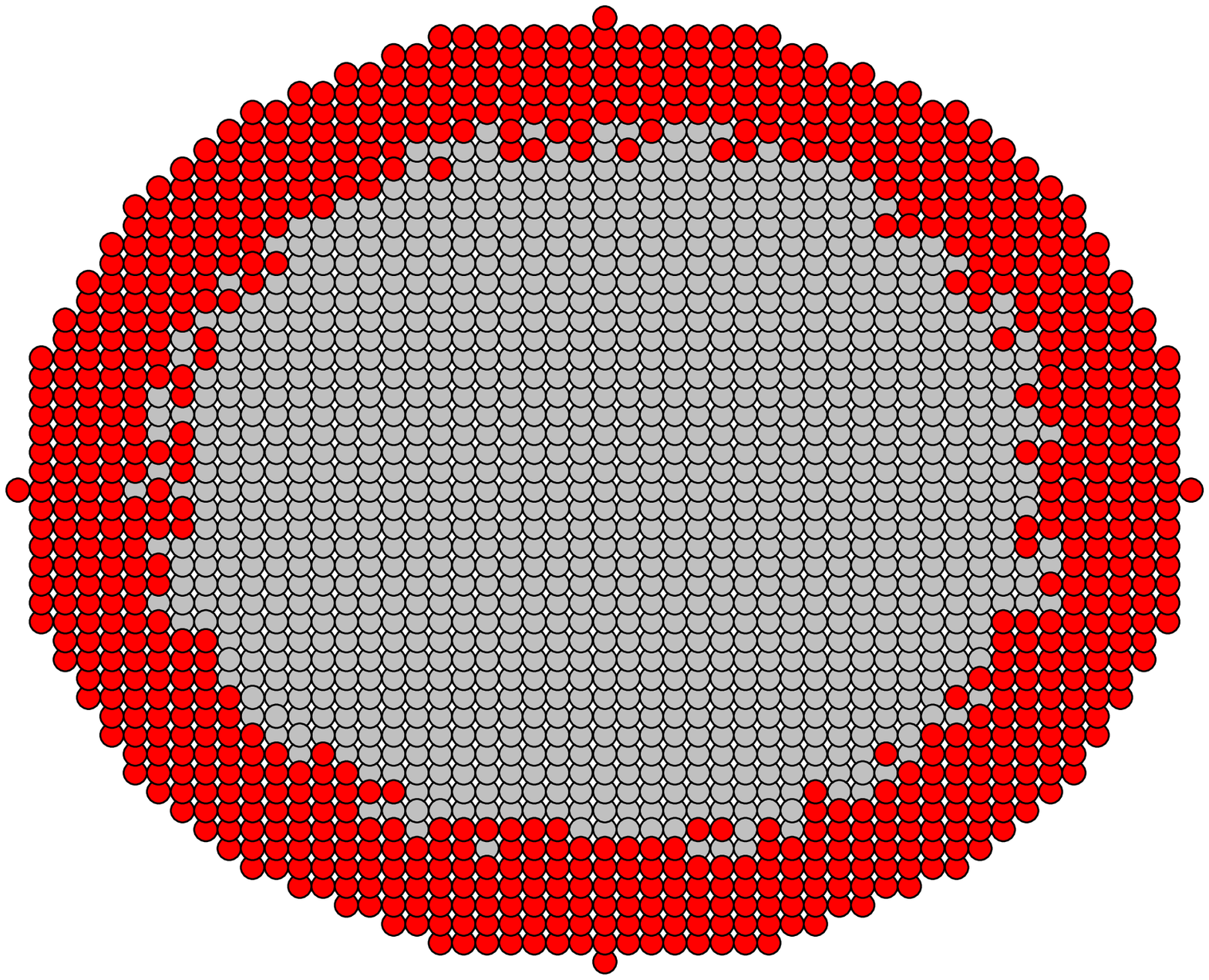}} \\
\subfigure[\hspace{0cm}]{\includegraphics[width=4.cm]{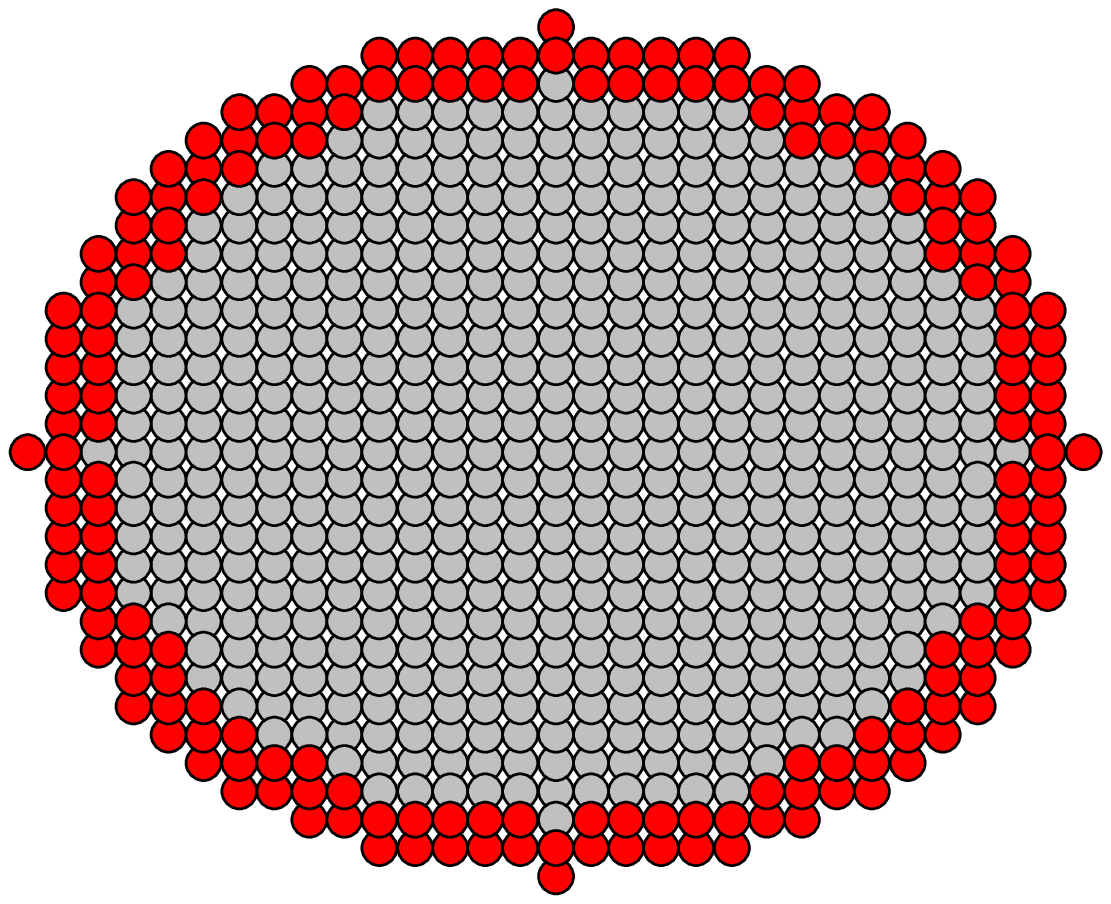}}
\subfigure[\hspace{0cm}]{\includegraphics[width=4.cm]{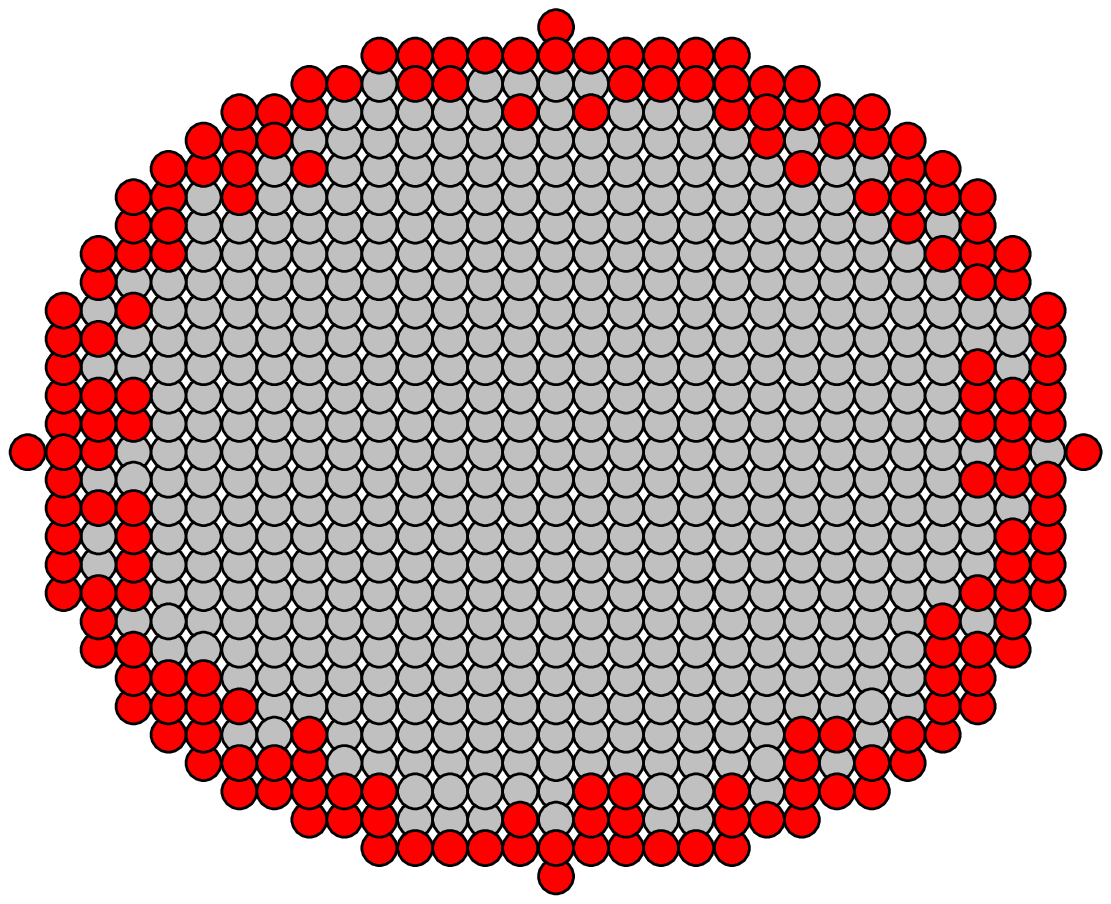}} \\
\subfigure[\hspace{0cm}]{\includegraphics[width=4.cm]{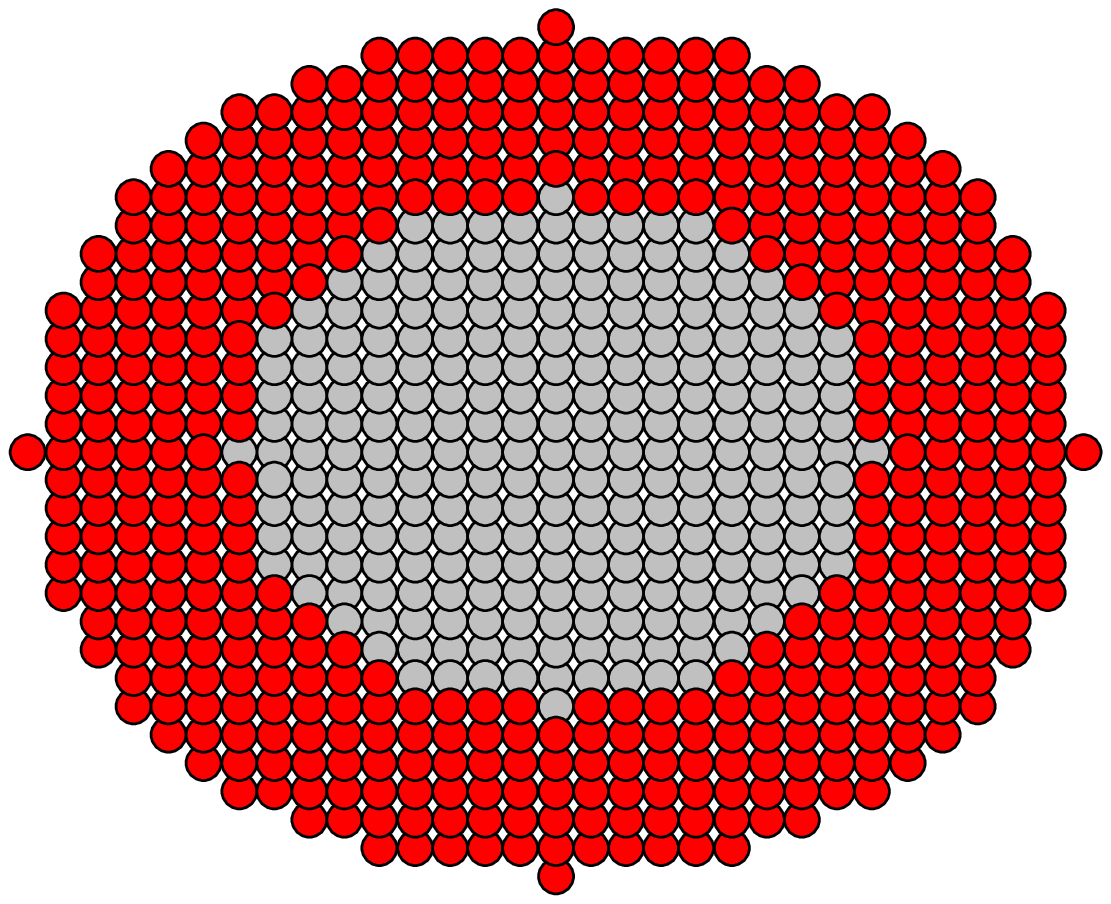}}
\subfigure[\hspace{0cm}]{\includegraphics[width=4.cm]{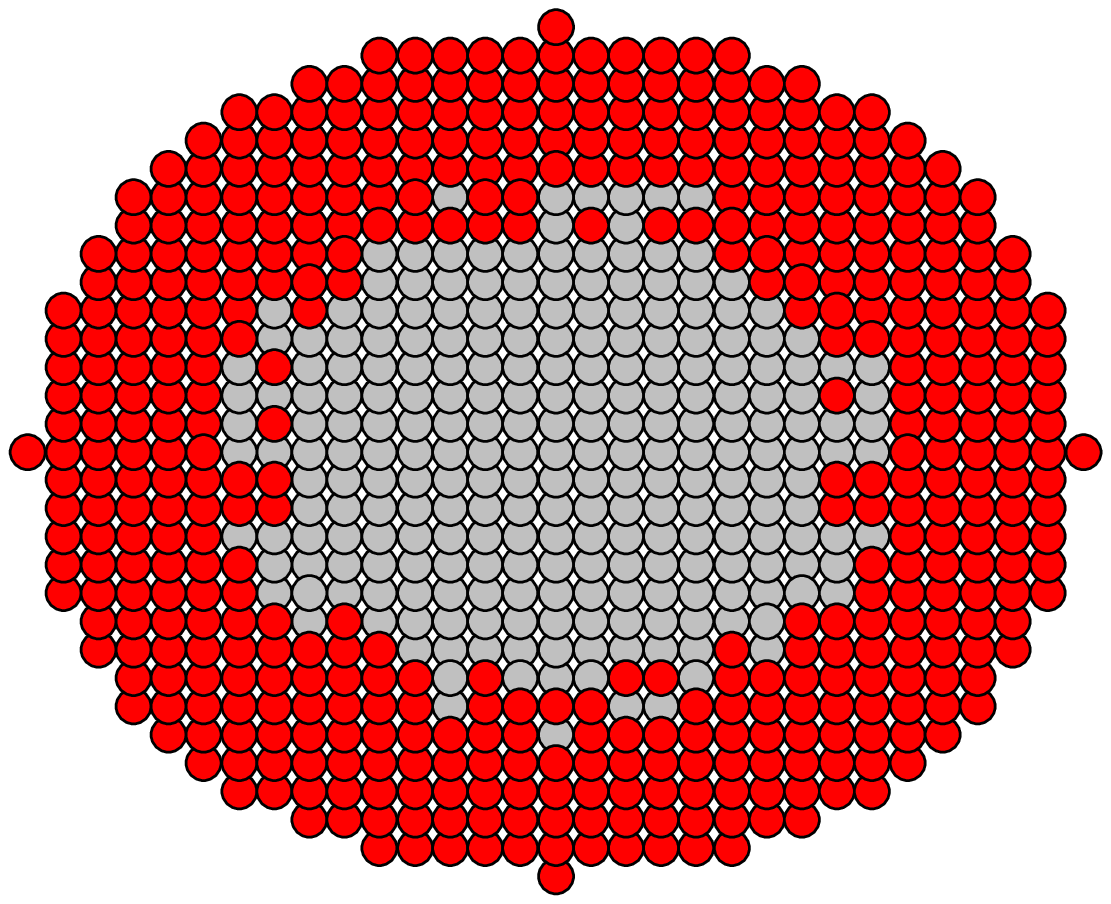}} \\
\caption{Schematic representation of 2D cross-sections of the simulated particles
with structural parameters enlisted in Table \ref{table1}.
The left and right columns show four different samples with or without roughness,
respectively. (a), (b): Sample 1; (c), (d): Sample 2; (e), (f): Sample 3; (g), (h): Sample 4.
$\sigma=1.0$ is used for generating the roughened particle interfaces. } \label{figx}
\end{figure*}

\begin{figure*}[!h]
\center
\subfigure[\hspace{0cm}] {\includegraphics[width=6cm]{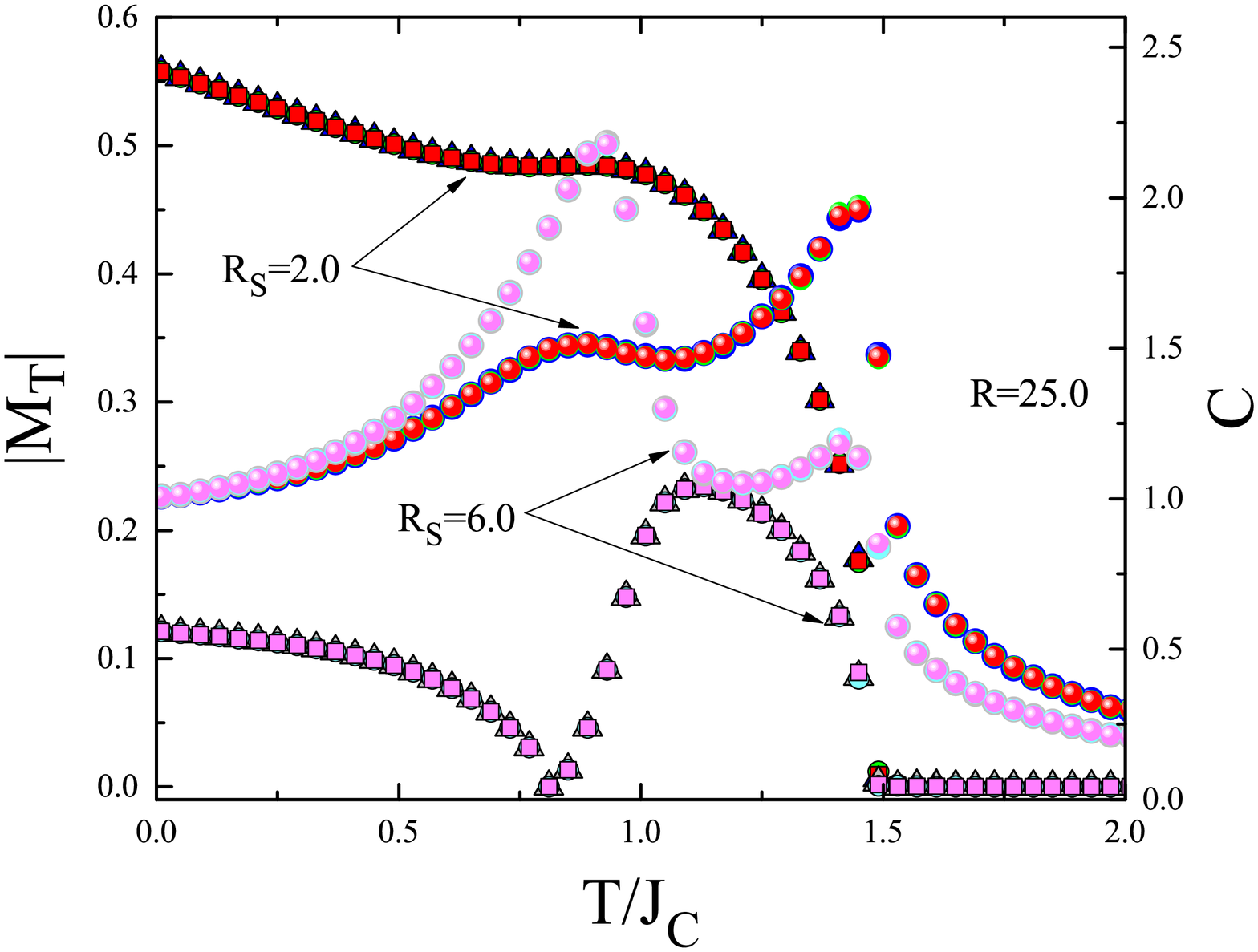}}
\subfigure[\hspace{0cm}] {\includegraphics[width=6cm]{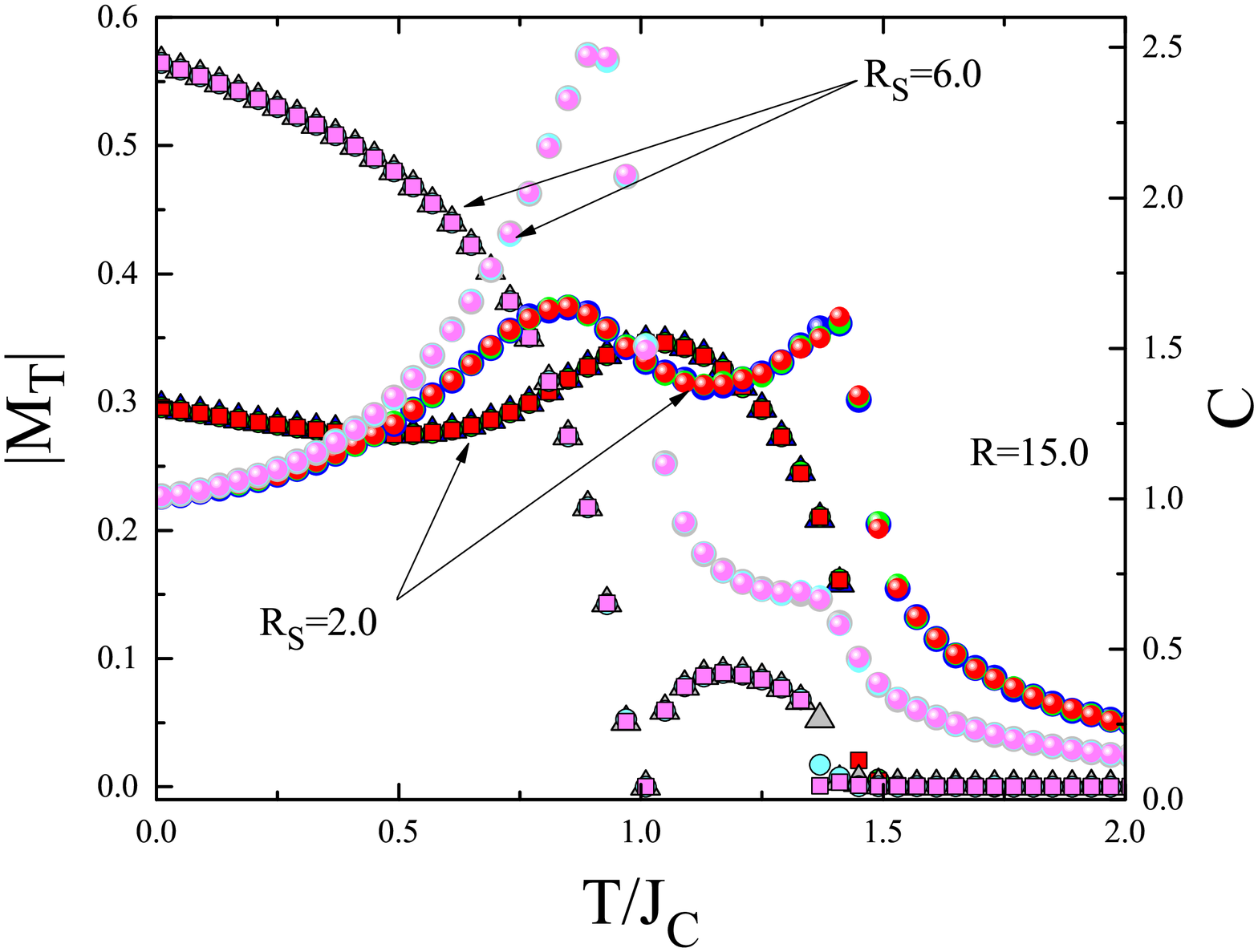}}\\
\caption{Temperature dependence of the total magnetization $|M_{T}|$ and
specific heat $C$ for the particles enlisted in Table \ref{table1}. The
system parameters are selected as $J_{S}=0.5J_{C}$, $J_{IF}=-0.5J_{C}$,
$K_{C}=0.1J_{C}$, $K_{S}=1.0J_{C}$. The curves corresponding to three
different degrees of roughness with $\sigma=0.01$, $0.5$ and $1.0$
overlap with each other.} \label{fig9}
\end{figure*}

\begin{figure*}[!h]
\center
\subfigure[\hspace{0cm}] {\includegraphics[width=6cm]{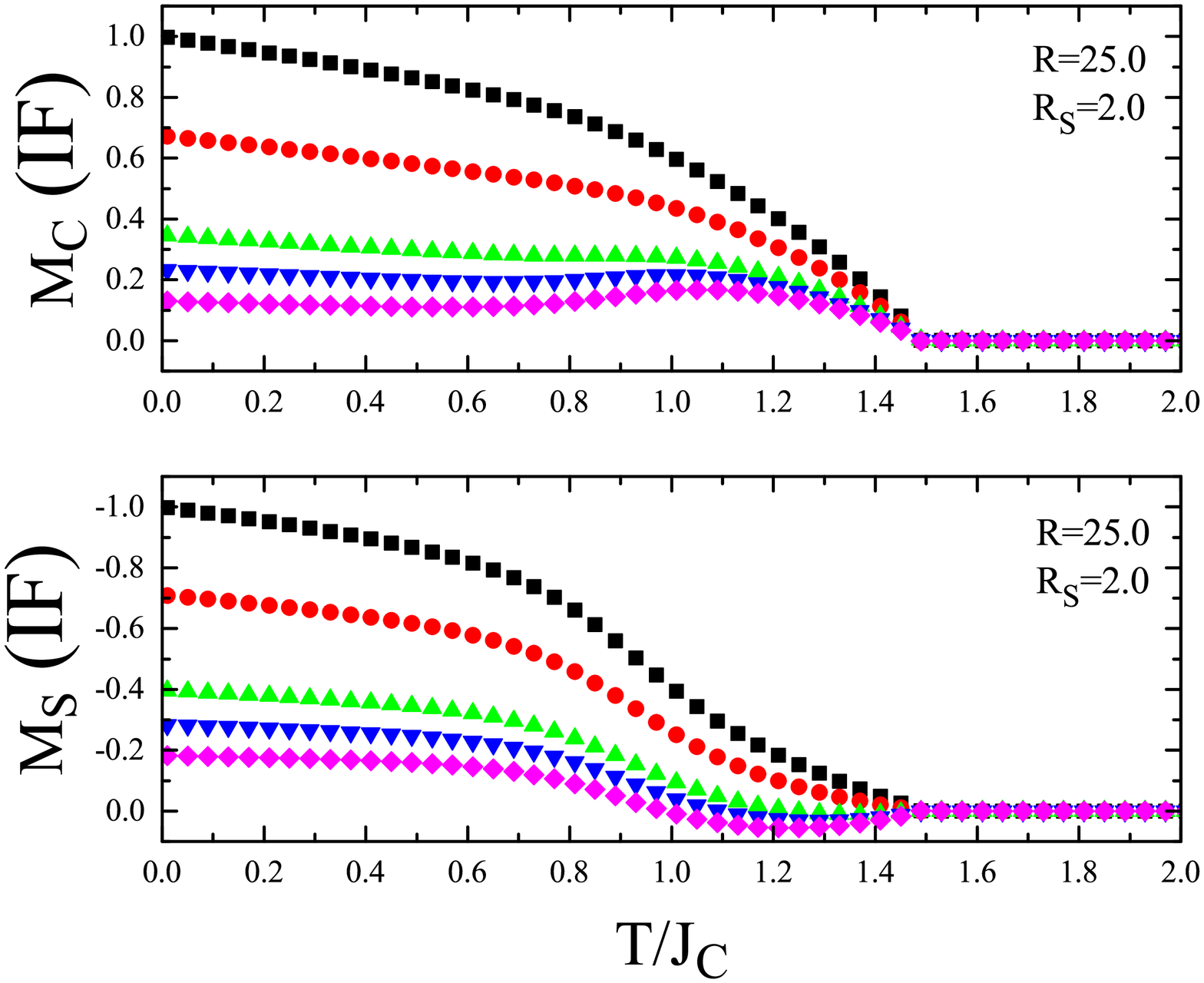}}
\subfigure[\hspace{0cm}] {\includegraphics[width=6cm]{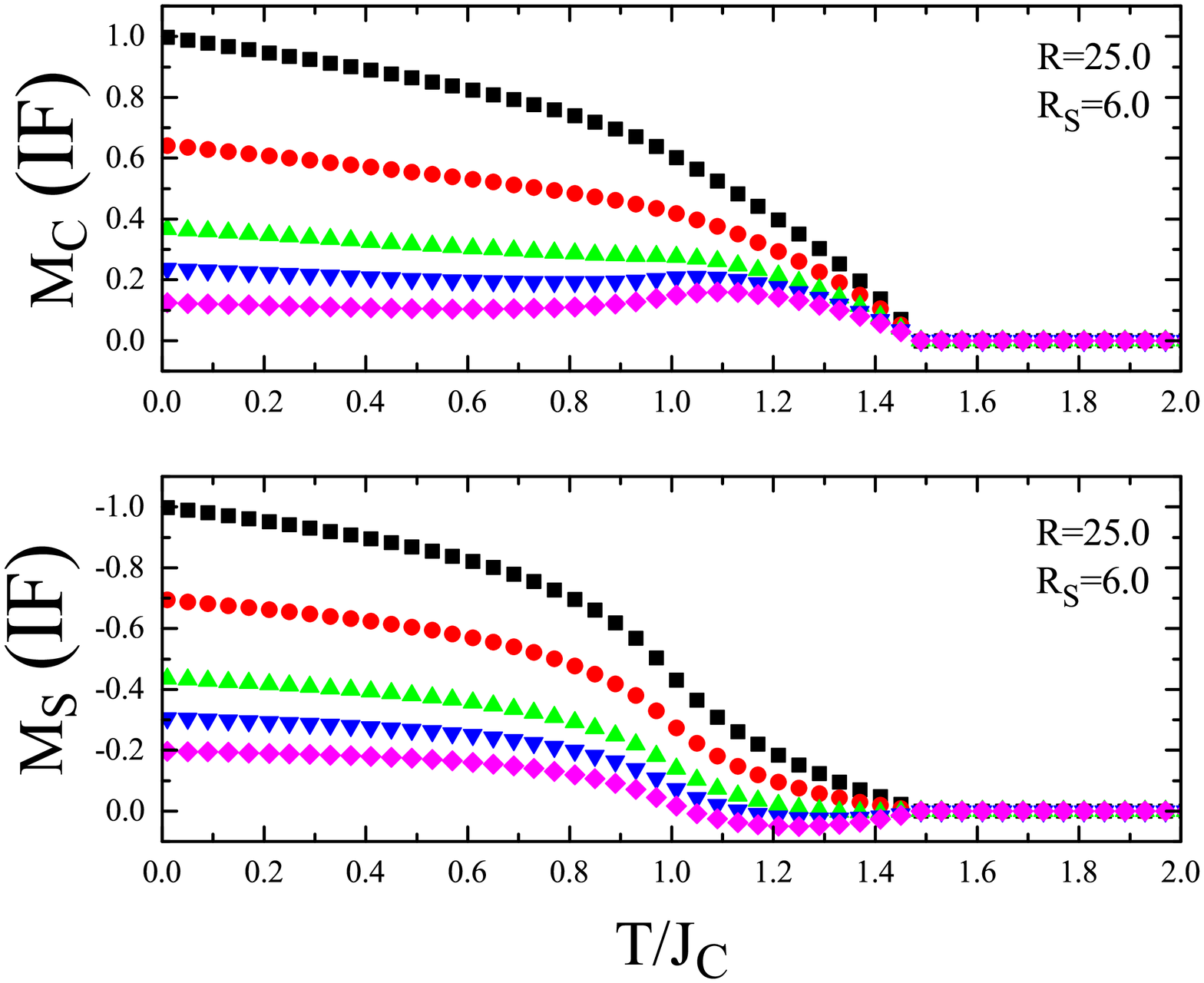}}\\
\subfigure[\hspace{0cm}] {\includegraphics[width=6cm]{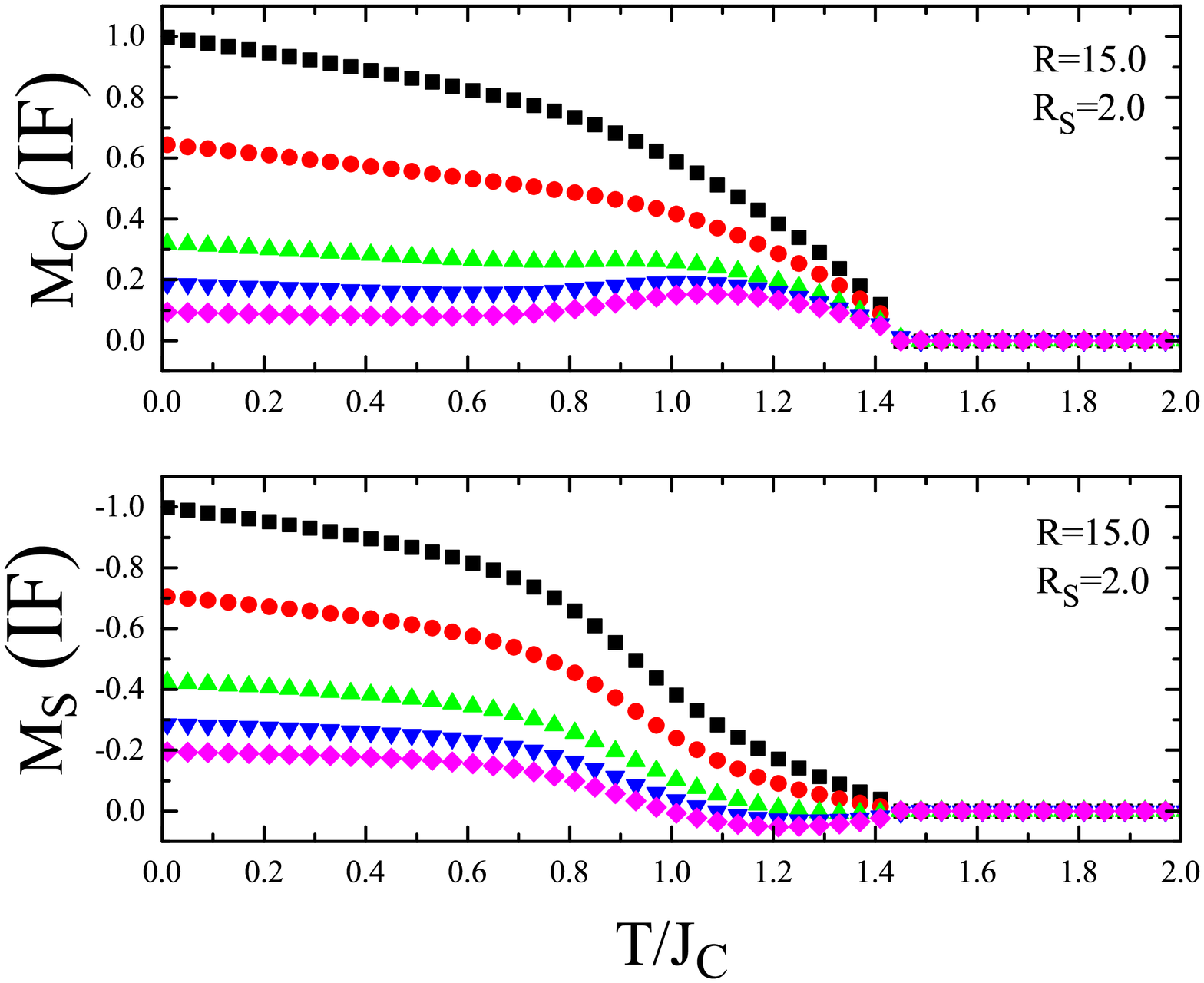}}
\subfigure[\hspace{0cm}] {\includegraphics[width=6cm]{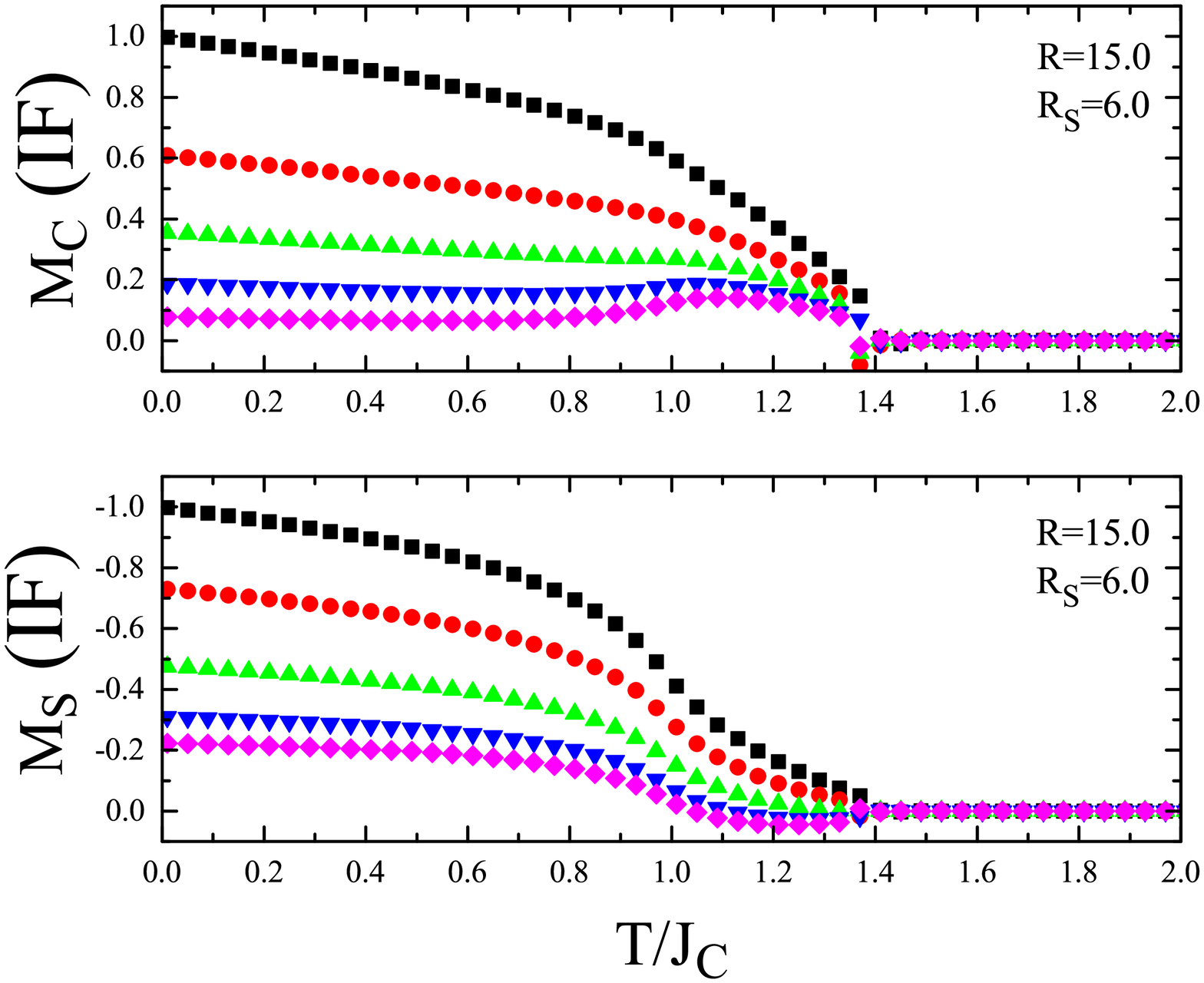}}\\
\caption{Thermal variation of the interface magnetization coming from  the core
and shell parts for particles prepared as (a) sample 1, (b) sample (2), (c)
sample 3, and (d) sample 4. Different symbols correspond to different
degrees of roughness such as $\sigma=0.01 (\blacksquare)$,
$\sigma=0.1 (\bullet)$, $\sigma=0.2 (\blacktriangle)$, $\sigma=0.3 (\blacktriangledown)$, $\sigma=0.5 (\blacklozenge)$.
The system parameters are as follows: $J_{S}=0.5J_{C}$, $J_{IF}=-0.5J_{C}$,
$K_{C}=0.1J_{C}$, $K_{S}=1.0J_{C}$. } \label{fig10}
\end{figure*}

Before proceeding further, it would be useful to clarify the effect of interface exchange
coupling $J_{IF}$ on the thermal and magnetic properties of the particle system. For this aim we
prepare sample realizations corresponding to four different samples (see Table \ref{table1}).
In Fig. \ref{fig7}, we show the influence of exchange coupling $J_{IF}$ on the thermal variation of
total magnetization $|M_{T}|$ and specific heat $C$ curves. At first sight, it is clear that
neither small nor large particles can exhibit a compensation point unless the shell thickness
is greater than a critical value (see Fig. \ref{fig2}a). Furthermore, even if the system exhibits
a compensation point $T_{comp}$, varying $J_{IF}$ values do not alter the location of $T_{comp}$.
A similar scenario is also valid for the transition temperature. However, varying $J_{IF}$ values
only affect the magnitude of the low temperature specific heat peaks. These aforementioned results may
indicate that the interfacial roughness may have no any significant effect on the compensation phenomenon.
Keeping this in mind, we prepare samples according to Table \ref{table1} with roughened interfaces. These structures
are also schematically illustrated in Fig. \ref{figx}. In Fig. \ref{fig9}, we depict the results
for $|M_{T}|$ and $C$ curves with four distinct samples corresponding to Fig. \ref{figx} from which one
can deduce that the degree of roughness does not play any significant role on the variation of both the
compensation point and critical temperature. This result supports our predictions based on Fig. \ref{fig7}.
Finally, let us show how the interface magnetizations regarding the core and shell regions are affected
from the roughness effects. In this context, Fig. \ref{fig10} concludes that increasing
amount of roughness does not alter the transition temperature of the system, but reduces
the low temperature saturation magnetizations of the core and shell interface regions.

\section{Concluding remarks}\label{conclude}
In conclusion, by benefiting from Monte Carlo simulation based on an 
improved Metropolis algorithm, we carry out a systematic 
investigation to elucidate the finite temperature magnetic properties of core-shell 
spherical nanoparticles  containing quenched surface and interface disorders as well as 
roughened interface effects. Some efforts have also been taken to study the 
particle size effects on  the thermal and magnetic properties of the particles. After a detailed 
numerical analysis, most prominent observations underlined in the present work 
can be briefly summarized as follows:

\begin{itemize}
 \item For core-shell spherical nanoparticles with imperfect surface layers,  it is 
 found that  bigger particles exhibit lower compensation point which decreases 
 gradually with increasing amount of vacancies, and vanishes at a 
 critical value. In view of the specific heat treatment as a function of the 
 temperature,  the specific heat demonstrates two successive peaks for clean case, 
 namely $p=1.0$.   However, as the amount of magnetic lattice sites in the shell progressively 
 decreases then the lower temperature peak tends to disappear.
 
 \item For particles with diluted  interface, our numerical results indicate that 
 there exists a region in the disorder spectrum where compensation temperature 
 linearly decreases with  decreasing dilution parameter.  It is interesting to note that 
 the linear variation region becomes  narrower for bigger particles.
 
 \item It is found that  the degree of roughness at the interface of the particle 
 does not play any  significant role on the variation of both the compensation  
point and critical temperature. However, the low temperature saturation 
magnetizations of the core and shell interface regions sensitively depend 
on the roughness parameter.
\end{itemize}

As a final conclusion, we note that more work is required to match and understand the theoretical 
and experimental findings  regarding the thermal and magnetic 
properties of the advanced functional core-shell nanoparticles in different 
geometries.

\section*{Acknowledgements}
The numerical calculations reported in this paper were performed
at TUBITAK ULAKBIM High Performance and Grid Computing Center (TR-Grid e-Infrastructure).


\begin{thebibliography}{10}
\bibitem{note2} Here, instead of critical temperatures observed in the thermodynamic limit,
we mean a pseudo critical one, since we are dealing with a finite particle.
\bibitem{note1} If a magnetic system exhibits compensation phenomenon then the spontaneous total magnetization reverses its sign
at a particular temperature called as the compensation temperature.
\end{thebibliography}
\end{document}